\documentclass[aps,pre,twocolumn,amsmath,amssymb,nofootinbib,floatfix,superscriptaddress]{revtex4-1}

\usepackage{amsmath,braket}
\usepackage{amssymb}
\usepackage{amsthm}
\usepackage[dvipdf]{color}
\usepackage{graphicx}
\usepackage{dcolumn}
\usepackage{bm,subfigure}
\usepackage{xcolor,colortbl}
\usepackage[normalem]{ulem}
\usepackage[OT1]{fontenc}

\usepackage{lineno}

\usepackage{amsmath,amssymb}

\usepackage{amsmath,amsfonts,amssymb,amsthm, bm}

\begin{document}

\title{
Edge-Response Renormalization Group for Multiscale Physical Networks
}

\author{Hao You}
\affiliation{Department of Physics, University of Michigan, Ann Arbor, MI 48109-1040, USA}
\affiliation{Center for Complex Particle Systems (COMPASS), University of Michigan, Ann Arbor, MI, USA.}

\author{Linlin Sun}
\affiliation{Center for Complex Particle Systems (COMPASS), University of Michigan, Ann Arbor, MI, USA.}
\affiliation{Department of Chemical Engineering, University of Michigan, Ann Arbor, Michigan 48109, USA}

\author{Shuaifeng Li}
\affiliation{Department of Physics, University of Michigan, Ann Arbor, MI 48109-1040, USA}
\affiliation{Center for Complex Particle Systems (COMPASS), University of Michigan, Ann Arbor, MI, USA.}

\author{Wenbing Wu}
\affiliation{Center for Complex Particle Systems (COMPASS), University of Michigan, Ann Arbor, MI, USA.}
\affiliation{Department of Chemical Engineering, University of Michigan, Ann Arbor, Michigan 48109, USA}
\affiliation{Michigan Center for Materials Characterization, University of Michigan, Ann Arbor, Michigan 48109, USA}

\author{Byeunggon Kim}
\affiliation{Center for Complex Particle Systems (COMPASS), University of Michigan, Ann Arbor, MI, USA.}
\affiliation{Department of Chemical and Biomolecular Engineering, North Carolina State University, Raleigh, NC 27695--7905, USA}

\author{Orlin D. Velev}
\affiliation{Center for Complex Particle Systems (COMPASS), University of Michigan, Ann Arbor, MI, USA.}
\affiliation{Department of Chemical and Biomolecular Engineering, North Carolina State University, Raleigh, NC 27695--7905, USA}

\author{Nicholas Kotov}
\affiliation{Center for Complex Particle Systems (COMPASS), University of Michigan, Ann Arbor, MI, USA.}
\affiliation{Department of Chemical Engineering, University of Michigan, Ann Arbor, Michigan 48109, USA}

\author{Xiaoming Mao}
\email{maox@umich.edu}
\affiliation{Department of Physics, University of Michigan, Ann Arbor, MI 48109-1040, USA}
\affiliation{Center for Complex Particle Systems (COMPASS), University of Michigan, Ann Arbor, MI, USA.}

\begin{abstract}
Renormalization provides a powerful route from microscopic interactions to emergent macroscopic behavior, but its extension to complex networks is obstructed by the absence of natural geometric scales. Here we introduce the edge-response renormalization group (ERRG), in which scale is defined by collective physical responses. ERRG identifies response-redundant degrees of freedom and organizes them into  supernodes, yielding a hierarchy of reduced network descriptions across scales. Applied to synthetic networks and experimentally characterized 2D nanowire and 3D aerogel  networks, ERRG captures both transport and mechanical responses, with low-frequency spectra collapsing under a single scale-dependent time rescaling. Importantly, the resulting renormalization flow reveals hidden, scale-dependent organization and provides a systematic way to track how effective network properties and characteristic dynamics evolve across scales.
\end{abstract}

\maketitle

\section*{The challenge of scale in physical networks} 

The success of the renormalization group (RG) in statistical physics rests on the homogeneity of the underlying space, which allows the same coarse-graining procedure to be applied iteratively, revealing flow of coupling coefficients and  
the notion of criticality
~\cite{kadanoff1966scaling,wilson1971renormalization}.

Many important materials in biology and emerging technologies are built from hierarchical network structures whose characteristics change across scales \cite{lakes1993materials,albert2002statistical,newman2003structure,cimini2019statistical}. Trabecular bone, for example, consists of mineralized collagen fibrils at nanometer scales, lamellae at micrometer scales, and porous trabecular networks at millimeter scales, giving rise to lightweight strength, damage tolerance, and load adaptation \cite{reznikov20153d}. Similar multiscale organization appears in transport networks such as leaf venation \cite{scarpella2024leaf}, hierarchical battery electrodes \cite{zhou2018recent,wang2022multifactorial}, electric circuits~\cite{huang2009impedance,nazerian2026frequency}, and brain networks \cite{betzel2017multi}. 
A unified approach to describe their multiscale properties will open the path to uniquely capable high-performance materials~\cite{mao2024complexity,jiang2020emergence}. 
Developing RG for these complex networks and characterizing the evolution of their dynamics across length scales remain an important open challenge \cite{villegas2023laplacian,gabrielli2025network,nurisso2025higher,cheng2021modelreduction}.

Recent approaches to network renormalization---focusing on complex network structures rather than physical responses---are revealing fascinating physics.
Box-covering methods provide an intuitive real-space coarse-graining procedure
~\cite{song2005self,song2006origins,goh2006skeleton,kim2007fractality,song2007calculate}; yet their reliance on shortest-path limited the applicability to heterogeneous small-world networks~\cite{villegas2023laplacian,gabrielli2025network}. Geometric renormalization overcomes this difficulty by embedding networks in a hidden metric space
~\cite{serrano2008self,boguna2021network,garcia2018multiscale,zheng2020geometric,zheng2024geometric}, yet its applicability depends on the existence and reliable inference of a meaningful latent geometry. 
Spectral approaches, such as Laplacian renormalization group (LRG) provide a dynamical definition of scale, preserving slow modes ~\cite{gfeller2007spectral,villegas2022laplacian,de2016spectral,ghavasieh2020statistical,ghavasieh2023generalized,ghavasieh2024diversity,de2017diffusion,harush2017dynamic,Schmidt2025,Dorfler2013}, yet these methods are defined  through diffusion and use global spectral computations, which limits their use for large networks with different physical processes.

Beyond structural renormalization of abstract graphs, \emph{networks as physical materials pose additional challenges and opportunities}. Different physical processes probe network structure in distinct ways: scalar transport such as electrical, thermal, or mass flow differs fundamentally from vector transmission such as mechanical stress, reflected for example in the separation between percolation and rigidity-percolation transitions \cite{stauffer2018introduction,feng1984percolation,jacobs1996generic}: A network may be sufficiently connected to support one class of response and yet still lack the structures required for another. Consequently, RG schemes for network materials must incorporate physics-specific rules to characterize scale-dependent responses
~\cite{broedersz2011criticality,mao2011coherent,mao2013elasticity,zhang2015rigidity,sharma2016strain,zhang2019correlated,rocklin2021elasticity,petridou2021rigidity,manning2024rigidity}.

Here we introduce ERRG for coarse graining of physical networks and revealing their multiscale organization, in a computationally efficient and experimentally feasible way. Through an edge-level response metric, which assesses how coherent the two endpoints are under the physical process of interest, 
ERRG constructs Kadanoff supernodes based directly on the physics of the network. 
We apply ERRG on both synthetic and experimentally characterized materials network data sets. 
Across ERRG steps $\ell$, 
we show that large-scale physical responses are well preserved, and the low-frequency spectra collapse upon a single time-rescaling factor $\tau^*_\ell$. The evolution of $\tau^*_\ell$ and network metrics following ERRG steps $\ell$ reveals \emph{scale-dependent hidden low-dimensional structures in these networks}. 
More broadly, ERRG establishes physical response as a basis for defining scale in complex physical networks, providing a route from network structure to emergent multiscale behavior.

\begin{figure*}[h!]
  \centering
  \includegraphics[width=0.9\textwidth]{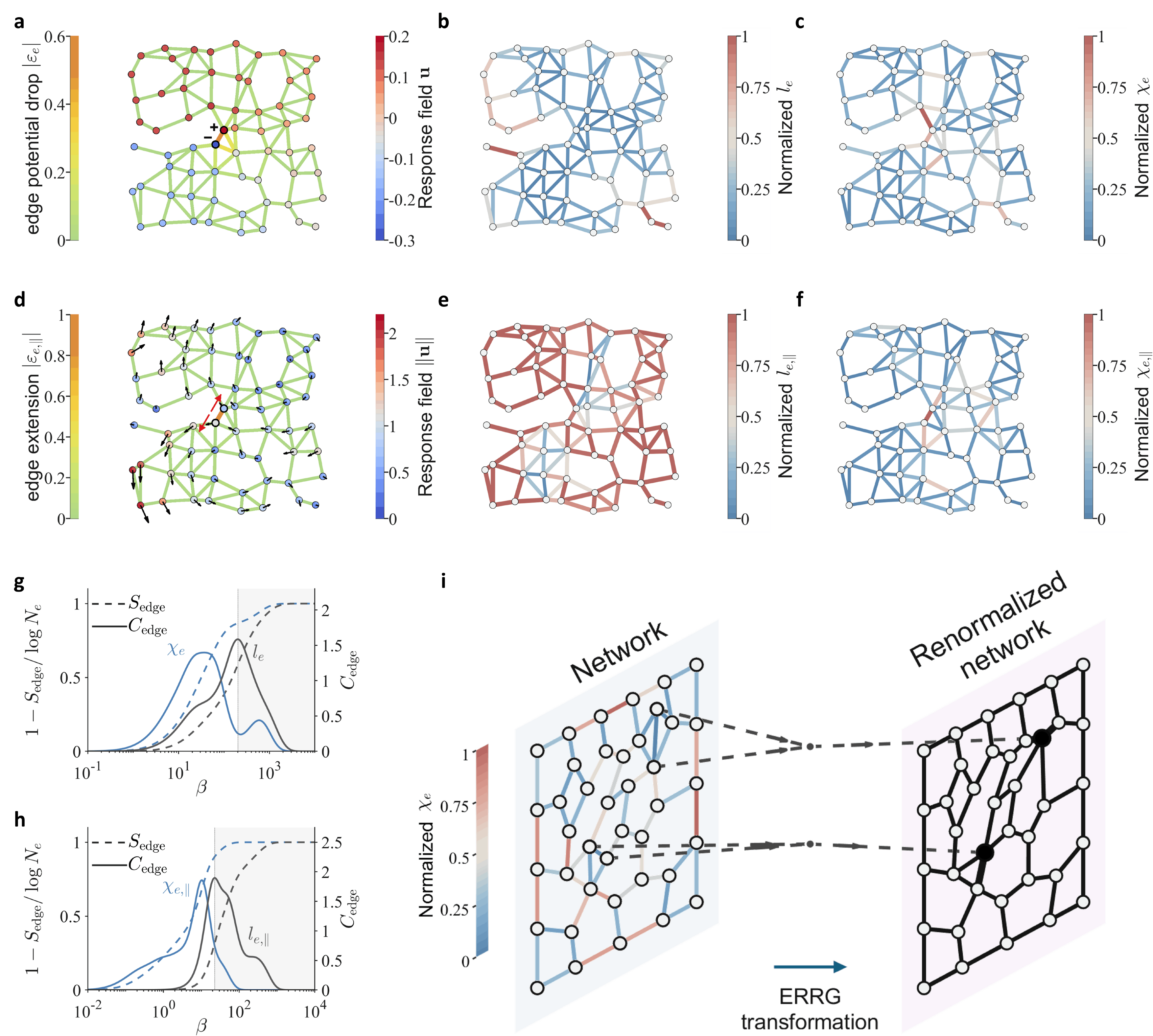}
    \caption{
    \textbf{Response-based edge scoring and supernode formation in ERRG.}
    \textbf{a,} A localized edge-dipolar excitation $(+,-)$ is applied across a selected edge. 
    Edge color denotes the magnitude of the induced static edge potential drop
    $|\varepsilon_e|$, and node color denotes the static node potential $\mathbf{u}$.
    \textbf{b,} The static edge-localized probe is repeated over all edges to compute the edge leverage score \(l_e\). 
    \textbf{c,} The edge-weight regularized response time edge score $\chi_e$, obtained through the dynamics of the same setting in (a). 
    \textbf{d--f,} Vector central-force response on the same network geometry, where a force dipole is applied along a chosen edge, leading to leverage score $l_{e,\parallel}$ in (e) and edge-weight-regularized response-time edge score $\chi_{e,\parallel}$ in (f) for the vector problem with longitudinal probe. Note that a large number of edges have $l_{e,\parallel}=1$ as they are \emph{isostatic} edges---removal of any causes zero modes. In contrast only a small number of edges in the scalar problem in (b) have \(l_e=1\)---removal of any causes isolated components. 
    \textbf{g,} Response-based edge entropy \(S_{\textrm{edge}}\)
    (normalized by \(\log N_e\), where \(N_e\) is the number of edges)
    and specific heat \(C_{\textrm{edge}}\) for the scalar edge scores
    \(\chi_e\) and \(l_e\).
    Peaks in \(C_{\textrm{edge}}\) identify crossover scales for response-based coarse graining.
    \textbf{h,} Response-based edge ensembles for the vector edge scores
    \(\chi_{e,\parallel}\) and \(l_{e,\parallel}\).
    \textbf{i,} Illustration of ERRG transformation.
    Connected components formed by low-\(\chi_e\) edges not selected in Step 1 are merged into supernodes to produce the
    renormalized network.
    }

  \label{Fig. 1}
\end{figure*}

\section*{Edge-response renormalization framework}
\label{Sec: theory}

\subsection*{Edge-response scores}
\label{Subsec:second_order_dynamics}

The guiding idea of ERRG is Wilsonian in spirit: coarse graining should eliminate microscopic degrees of freedom (DOFs) while preserving their influence in an effective theory at larger scales
~\cite{wilson1971renormalization,wilson1974renormalization}.
In conventional RG, this organization is primarily by length scale: short-wavelength modes are integrated out, and their influence appears through renormalized couplings in the long-wavelength theory. For physical networks of complex structures, due to the lack of clear association of DOFs with length scales as discussed above, we instead identify \emph{important DOFs through edge-based responses}.

Although the idea of ERRG is general in principle, here we first develop a concrete realization for linear reciprocal networks. For such systems, the driven response problem can be described as
\begin{equation}
\left[\mathbf A(\partial_t)+\mathbf L\right]\mathbf u(t)
=
{\mathbf f}(t),
\label{Eq:response_real_time}
\end{equation}
where $\mathbf{u}(t),\mathbf{f}(t)$ are generalized node potential responses (e.g., displacement, electric potential and  temperature), and external loading (e.g., force, injected electric and heat current). 
The local response operator $\mathbf A(\partial_t)$ collects node dynamical or local environmental terms: 
For overdamped relaxation, \(\mathbf A(\partial_t)=\boldsymbol{\Gamma}\partial_t\);
For damped second-order dynamics, \(\mathbf A(\partial_t)=\mathbf M \partial^2_t+ \boldsymbol{\Gamma} \partial_t\).
The operator \(\mathbf L\) encodes network-mediated couplings, 
$\mathbf L=\mathbf C^T\mathbf K\mathbf C $,
where \(\mathbf C\) maps (scalar or vector) nodal DOFs to edge DOFs and \(\mathbf K\) is a diagonal matrix of generalized edge weights $w_e$ (e.g., spring constants, conductance for electric and heat currents)~\cite{strang1986introduction,pellegrino1986matrix,mao2018maxwell}.

This problem can be generally solved via Laplace transform, 
\begin{equation}
\left[\mathbf{\hat{A}}(s)+\mathbf L\right]\widehat{\mathbf u}(s)
=
\widehat{\mathbf f}(s),
\qquad
\mathbf G(s)
=
\left[\mathbf{\hat{A}} (s)+\mathbf L\right]^+ .
\label{Eq:general_response_operator}
\end{equation}
Here \(s\) is the Laplace variable, \(\widehat{\mathbf u}(s)\) and \(\widehat{\mathbf f}(s)\) are the transformed nodal response and external loading vectors, respectively, and \((\cdot)^+\) denotes the Moore--Penrose pseudoinverse \cite{ben2003generalized}. The Green's operator \(\mathbf G(s)\) 
maps the external loading to the nodal response in Laplace space \cite{kubo1957statistical}.

An important measure we build our RG framework on is the \emph{edge-level response}, which answers the question of whether two nodes connected by an edge should be merged as a supernode. 
Imagine an experiment of suddenly imposing a force dipole on an edge in a mechanical network. 
The response propagates through the network over time. The resulting \emph{spatiotemporal} pattern of displacements offers a comprehensive probe of the network response. 

This is captured by the projection of the Green's function onto each edge $e$, $\mathcal{G}_e(s)\equiv\mathbf b_e^\top\mathbf G(s)\mathbf b_e$, where $\mathbf b_e$ is the edge incidence vector (see Supplementary Information Section~\ref{SI:edge_projection_vector}). 
The first two moments of $\mathcal{G}_e(s)$ are 
\begin{equation}
    R_e
=\mathcal{G}_e(0), \quad \tau_e
=
-\left.
\frac{\partial}{\partial s}
\log \mathcal{G}_e(s)
\right|_{s=0} .
\end{equation}
Here $R_e$ is the effective resistance~\cite{spielman2008graph}, and relates to the 
edge leverage score as $l_e=w_eR_e$~\cite{drineas2010effective}.
Meanwhile, $\tau_e$ measures the relaxation time of dipole signals injected on  edge $e$, and becomes large when the edge is dominated by \emph{slow modes}. In the framework of RG, slow modes are essential for obtaining effective theories.
These responses can be conveniently implemented numerically or experimentally with low time complexity, through shifted steady-state measurements (see Supplementary Information Section~\ref{SI:shifted_steady_response}).

The formulation described above is general for any physical networks, but we now specialize to overdamped (first-order in time) dynamics to offer a concrete demonstration. In this case, $\mathbf{\hat{A}}(s)=\boldsymbol{\Gamma}s$ and $\tau_e=\frac{\mathbf b_e^\top\mathbf L^+\mathbf{\Gamma}\mathbf L^+\mathbf b_e}{ \mathbf b_e^\top\mathbf L^+\mathbf b_e}$. %
Interestingly, for scalar networks with unit nodal drag, this coincides with an optimization-derived edge-contraction threshold in spectral graph coarsening~\cite{bravo2019unifying}. We define the edge-weight-regularized response time
\begin{equation}
\chi_e=
w_e\tau_e ,
\label{Eq:ERRG_score}
\end{equation}
to be the main edge score $E_e$ to contract edges  in ERRG, $E_e=\chi_e$. 
This score favors edges that are \emph{both strongly weighted and involved in slow response pathways}. Different edge scores $E_e$, such as $E_e=l_e$ can also be chosen to run ERRG (Supplementary Information Section~\ref{SI:nanowire_method_comparison}).

\subsection*{Implementation of ERRG via randomized sampling}
\label{Subsec:ERRG_implementation}

According to the chosen edge score $E_e$, the ERRG procedure runs as follows for each iteration $\ell$ (Fig.~\ref{Fig. 1}i and Supplementary Information Section~\ref{SI:response_biased_sampling}).

\begin{enumerate}
\item \textbf{Sample edges.}
Compute edge scores $E_e$ for all edges at iteration $\ell$. 
Sample \(q_\ell\) edges with replacement, with per-draw probability for edge $e$ being \(p_e=E_e/\sum_{e'}E_{e'}\). \(q_\ell\) is determined by the sampling parameter \(\varepsilon_{\mathrm{RG}}\), chosen at the beginning. 

\item \textbf{Construct contraction blocks.} Endpoints of the low-\(E_e\) edges not selected in Step 1 are contracted into supernodes.

\item \textbf{Build the coarse graph.}
The coarse graph is constructed by Kadanoff supernode projection, which aggregates the corresponding node weights and parallel edge weights.

\end{enumerate}

This procedure was inspired by the sampling-budget scaling of spectral sparsification \cite{spielman2008graph,spielman2004nearly}, but with an important distinction that we merge edges to obtain supernodes, instead of removing them as in spectral sparsification. 
Iterating the steps above produces a hierarchy
\[
\mathbf{L}^{(\ell=0)}
\rightarrow
\mathbf{L}^{(\ell=1)}
\rightarrow
\mathbf{L}^{(\ell=2)}
\rightarrow
\cdots .
\]
The induced renormalization flow is indexed by the RG level \(\ell\). At each step, the probability for an edge to be contracted scales as \((1-p_e)^{q_\ell}\simeq\exp(-q_\ell p_e)\), so low-\(E_e\) edges are exponentially more likely to be contracted. 

This procedure naturally defines an edge score  ensemble,  
\begin{equation}
Z_{\textrm{edge}}(\beta)=\sum_e e^{-\beta E_e},
\qquad
\mu_e(\beta)=\frac{e^{-\beta E_e}}{Z_{\textrm{edge}}(\beta)} ,
\label{Eq:edge_distribution}
\end{equation}
where the response-resolution parameter \(\beta\) controls how selective the procedure is.
The corresponding Gibbs--Shannon entropy of this ensemble is then 
$
S_{\textrm{edge}}(\beta)
=
-\sum_e \mu_e(\beta)\log \mu_e(\beta), 
$
defining the corresponding specific heat 
$C_{\textrm{edge}}(\beta)
=
-\beta\frac{dS_{\textrm{edge}}}{d\beta}=\beta^2
\operatorname{Var}_\beta(E)$.
Peaks in \(C_{\textrm{edge}}(\beta)\) indicate values of the response-resolution parameter $\beta$ at which the sampling hierarchy changes most strongly. 

\section*{ERRG reveals multiscale structures in physical networks}
\subsection*{Synthetic multiscale physical networks}

To test the capability of ERRG in revealing underlying structures in complex physical networks, we create a class of intrinsically multiscale physical networks, inspired by self-assembled gel networks. These networks consist of dense clusters of nano- or colloidal particles at the scale of $nm \sim \mu m$ but porous networks at larger scales from kinetically arrested phase separation~\cite{zaccarelli2007colloidal}. Thus they exhibit responses that combine different physics across scales.
We create these multiscale networks by 
first generating a Voronoi skeleton as a large-scale backbone, and then attaching nodes around the backbone, using the protocol described in Supplementary Information Section~\ref{SI:voronoi_ensemble}. 
All edges are assigned uniform weight, so that the response heterogeneity arises entirely from network connectivity. 

\begin{figure*}[t]
    \centering
    \includegraphics[width=\textwidth]{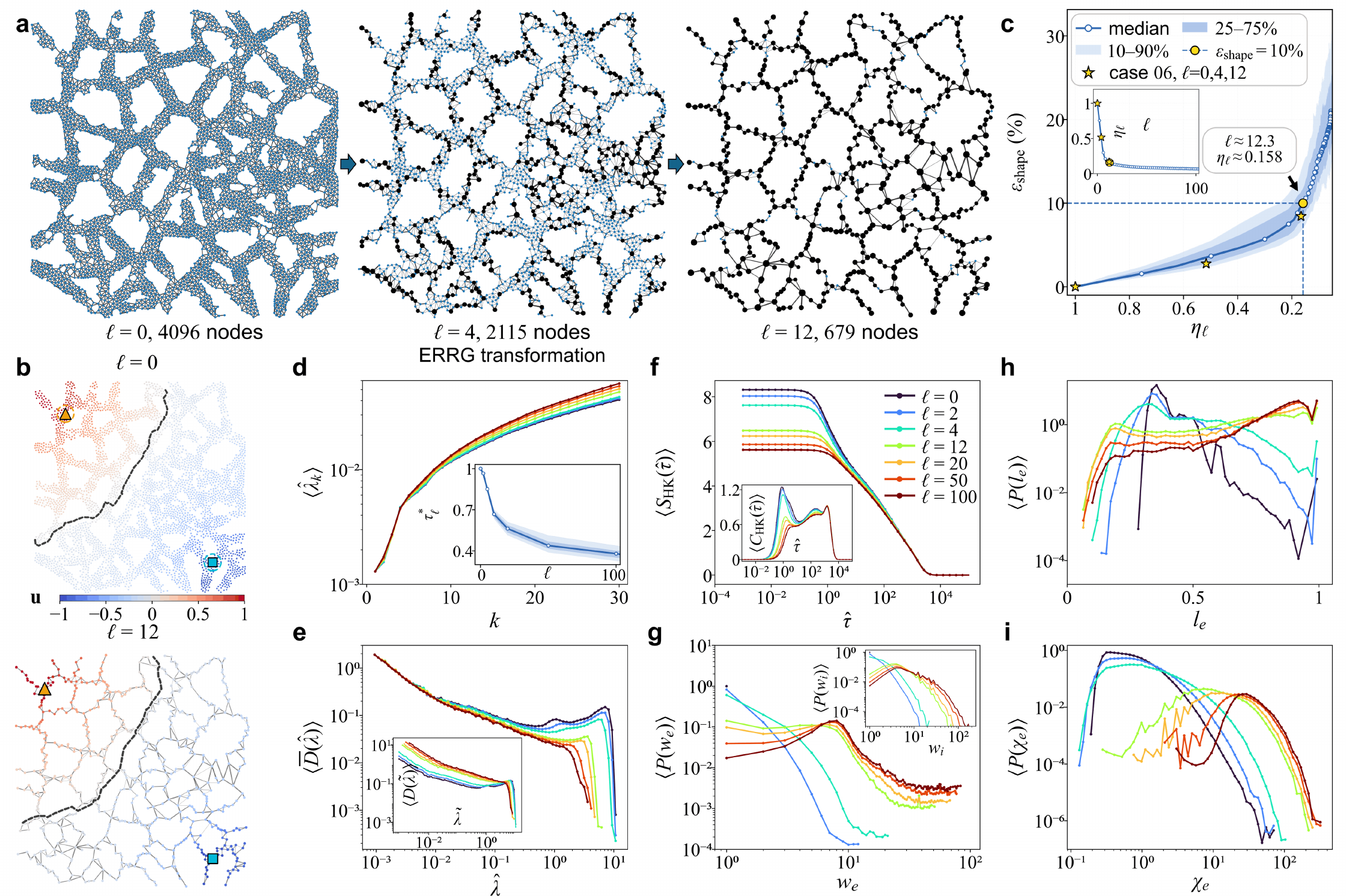}
    \caption{
    \textbf{ERRG flow of multiscale physical networks.}
    \textbf{a,} Representative ERRG trajectory on a multiscale physical network, showing substantial node reduction while retaining the large-scale structure.
    \textbf{b,} Static-response preservation under two-patch \(\pm\) loadings. Colors denote the discrete scalar response field \(\mathbf u\), and the dashed line marks the contour $\mathbf{u}=0$. 
    \textbf{c,} Ensemble error--compression tradeoff over 90 independent network realizations. 
    Stars mark the representative levels in  (a,b). The inset shows compression over ERRG iterations $\ell$. 
    \textbf{d,} The  ensemble-averaged rescaled lowest 30 eigenvalues \(\langle\widehat{\lambda}_k\rangle\); The inset shows the ensemble-median time-rescaling factor \(\tau_\ell^*\).
    \textbf{e,} The  mode-count-corrected spectral density \(\langle\bar D(\widehat{\lambda})\rangle\), obtained by weighting the normalized spectral density by the retained fraction of nonzero Laplacian modes. The inset shows the unrescaled accumulated-node-weight-normalized spectral density \(\langle D(\widetilde{\lambda})\rangle\).
    \textbf{f,} Heat-kernel entropy \(\langle S_{\mathrm{HK}}(\widehat{\tau})\rangle\), which quantifies how broadly the heat-kernel weight is distributed among Laplacian modes. The inset shows the heat-kernel specific heat \(\langle C_{\mathrm{HK}}(\widehat{\tau})\rangle\). 
    \textbf{g,} Ensemble-averaged edge weight distribution \(\langle P(w_e)\rangle\) at selected RG levels. {The inset shows the corresponding ensemble-averaged accumulated node weight distribution \(\langle P(w_i)\rangle\).}
    \textbf{h,} Ensemble-averaged leverage score distribution
    \(\langle P(l_e)\rangle\). 
    \textbf{i,} Ensemble-averaged edge-weight regularized response time distribution
    \(\langle P(\chi_e)\rangle\).
    Colors in (d--i) follow the legend in (f).
    All ERRG trajectories use the fixed sampling parameter \(\varepsilon_{\mathrm{RG}}=2.8\).
    }
    \label{Fig. 2}
\end{figure*}

We apply ERRG to these multiscale networks, with a fixed sampling parameter \(\varepsilon_{\mathrm{RG}}=2.8\). %
{This value is selected from the ensemble analysis by identifying the main \(C_{\textrm{edge}}(\beta)\) peak and considering the error--compression tradeoff on its low-\(\beta\) side, as detailed in Supplementary Information Section~\ref{SI:voronoi_ensemble}.}
Figure~\ref{Fig. 2}a shows a representative network and demonstrates that
ERRG can achieve substantial compression within only a few iterations.
The successive coarse-grained representations retain and progressively reveal the large-scale backbone of the original network. Although all edges have uniform weight initially, variations in geometry and connectivity generate a heterogeneous spatial distribution of \(E_e\), indicating that the edge score encodes the multiscale organization of the network. Low-score edges tend to occur within locally dense and redundant regions, whereas high-score edges trace large-scale pathways that support the slow physical response. 

ERRG substantially reduces the network size while retaining the large-scale backbone.
We quantify compression at RG level \(\ell\) by the
node-count ratio, \(\eta_\ell=N_\ell/N_0\), where \(N_\ell\)
and \(N_0\) denote the numbers of nodes at level \(\ell\) and in the
original network, respectively. 

To demonstrate the preservation of large-scale physical responses, we perform a two-patch \(\pm\) loading test (Fig.~\ref{Fig. 2}b), and quantify the preservation of the entire response field by a shape-response error \(\varepsilon_{\mathrm{shape}}\) by lifting the coarse graph to the same number of nodes as the original but carrying the response of the coarse graph (defined in Supplementary Information Section~\ref{SI:response_error}). The ensemble statistics of this test show that 
the ensemble-median shape-response error \(\varepsilon_{\mathrm{shape}}\) remains below \(10\%\) up to an interpolated RG level of \(\ell\approx12.3\), corresponding to a node-count ratio of \(\eta_\ell\approx0.158\), for this type of network (Fig.~\ref{Fig. 2}c). Beyond this point, the increase in \(\varepsilon_{\mathrm{shape}}\) indicates that further contraction begins to remove pathways that contribute appreciably to the static response.

We also quantify ensemble-level spectral diagnostics during ERRG (Fig.~\ref{Fig. 2}(d--e)). We show that the Laplacian eigenvalues are well-preserved: the bare Laplacian obtained in ERRG by summing edge weights during coarse graining shows preserved eigenvalue structures but an overall trend of shifting all eigenvalues up, a phenomenon known in network coarse graining \cite{villegas2023laplacian}. This is partially (but not completely) removed by using the accumulated-node-weight normalization of the Laplacian~\cite{CraigBampton1968,Koren2003}, 
\begin{equation}\label{EQ:drag_normal}
\widetilde{\mathbf L}^{(\ell)}
=
\left(\boldsymbol{W}^{(\ell)}\right)^{-1/2}
\mathbf L^{(\ell)}
\left(\boldsymbol{W}^{(\ell)}\right)^{-1/2},
\end{equation}
where $\boldsymbol{W}^{(\ell)}=\boldsymbol{\Gamma}^{(\ell)}$ for overdamped dynamics and $\boldsymbol{W}^{(\ell)}=\boldsymbol{M}^{(\ell)}$ for inertial dynamics.  
We find that an extra rescaling factor 
{\(\tau_\ell^*=
\widetilde{\lambda}_1^{(0)}/\widetilde{\lambda}_1^{(\ell)}\) can be introduced, 
where \(\widetilde{\lambda}_1\) is the first nonzero
eigenvalue. This defines the rescaled eigenvalues as 
\(\widehat{\lambda}_k^{(\ell)}
=\tau_\ell^*\widetilde{\lambda}_k^{(\ell)}\). 
Remarkably, this rescaling approximately collapses the eigenvalues and their density of states across
RG levels onto one master curve, demonstrating excellent preservation of the large-scale network dynamics (Fig.~\ref{Fig. 2}d,e). } 

This preservation of dynamics also shows up in other measures, 
including the heat-kernel entropy
\(S_{\mathrm{HK}}(\widehat{\tau})
=-\sum_{k\ge1}p_{k,\mathrm{HK}}(\widehat{\tau})
\log p_{k,\mathrm{HK}}(\widehat{\tau})\)
and heat-kernel specific heat
\(C_{\mathrm{HK}}(\widehat{\tau})
=-\widehat{\tau}\,
dS_{\mathrm{HK}}(\widehat{\tau})/d\widehat{\tau}\),
where
\(p_{k,\mathrm{HK}}(\widehat{\tau})
=
e^{-\widehat{\tau}\widehat{\lambda}_k}/
\sum_{j\ge1}e^{-\widehat{\tau}\widehat{\lambda}_j}\).
They characterize the diversity of active Laplacian modes and the
crossovers between spectral scales, respectively
\cite{villegas2023laplacian}. As seen in Fig.~\ref{Fig. 2}f, they are preserved for long rescaled times $\widehat{\tau}$, consistent with the idea of slow-mode preservation of ERRG. 

Interestingly, although preserving large-scale response, ERRG evolves local properties of the network across RG levels. As shown in Fig.~\ref{Fig. 2}(g--i), the distributions of edge and node weights ($P(w_e), P(w_i)$), although starting as uniform across all edges and nodes, shift to develop heavy tails characterizing super edges and supernodes. This is accompanied by the evolution of edge leverage score $l_e$, which flattens over the range of $[0,1]$, demonstrating \emph{growing heterogeneity} and emergence of bottlenecks; and edge-weight regularized response time $\chi_e$ (our primary score used in ERRG), which shifts to larger values, consistent with the ERRG scheme of maintaining slow modes.

\subsection*{Central-force mechanical networks}

\begin{figure*}[t]
    \centering
    \includegraphics[width=\textwidth]{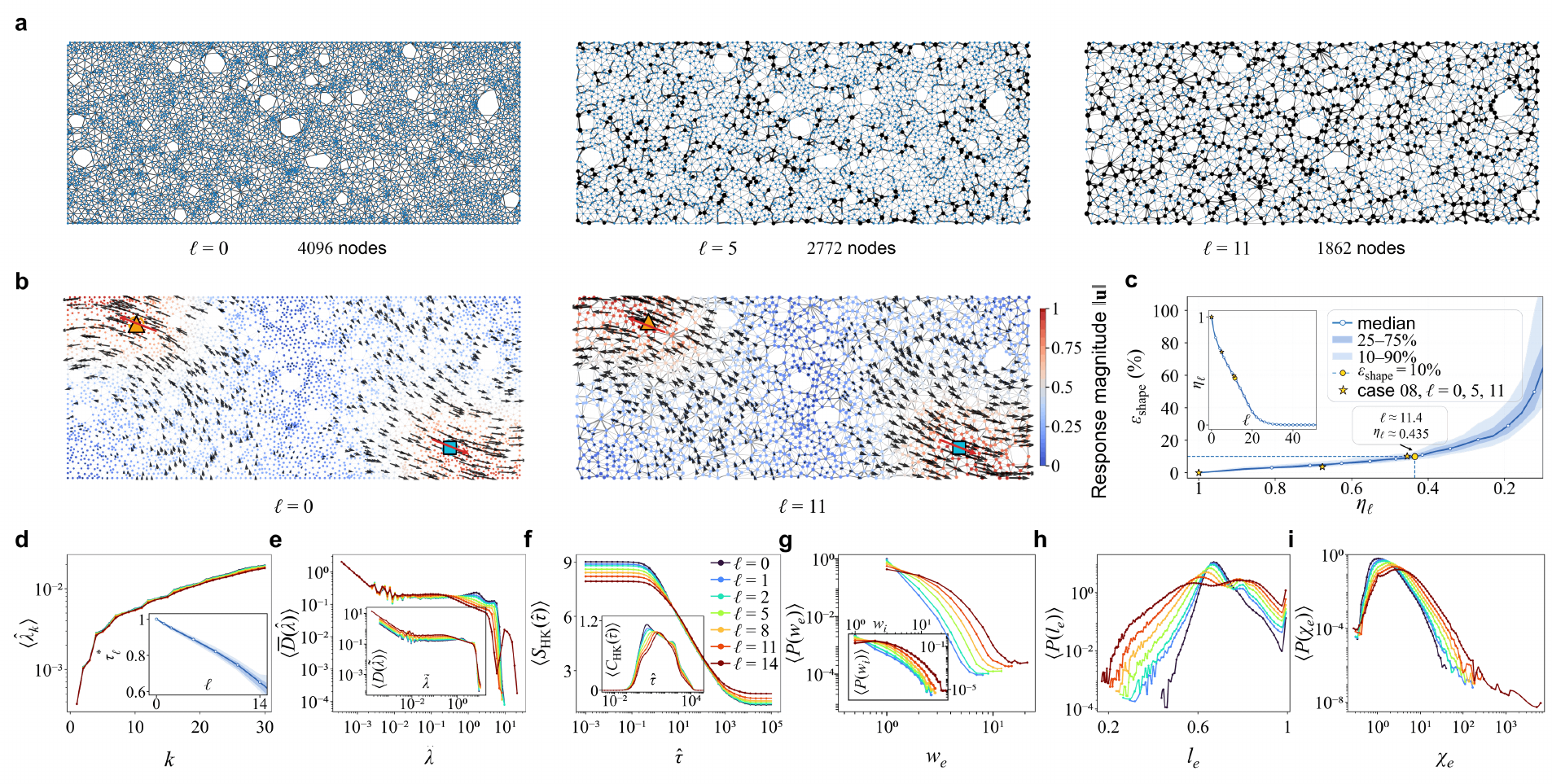}
    \caption{
    \textbf{ERRG flow of vector mechanical networks with multiscale holes.}
    \textbf{a,} Representative ERRG trajectory for a central-force
    mechanical network generated from a Delaunay
    triangulation with multiscale holes. 
    \textbf{b,} Vector-response preservation under prescribed displacements applied over two loading patches. Colors denote the normalized displacement magnitude $\|\mathbf{u}\|$, red arrows denote the prescribed patch displacements, and black arrows show the resulting displacement response.
    \textbf{c,} Ensemble error--compression tradeoff over 90 independent realizations.
    \textbf{d--i,} Ensemble-averaged network measures (median for \(\tau_\ell^*\)) across RG steps $\ell$, including rescaled low-order eigenvalues and time-rescaling factor (d), mode-count-corrected spectral density (e), heat-kernel entropy and specific heat (f), edge weight and node weight distributions (g), edge leverage score distribution (h), and edge-weight regularized response time distribution (i). 
    In panels ({{d--f}}), the same drag normalization and time
    rescaling as in Fig.~\ref{Fig. 2} are applied to the (mechanical) $\mathbf L^{(\ell)}$. In these spectral and heat-kernel diagnostics, only the
    prescribed rigid-body modes are removed; additional zero or
    near-zero modes generated during coarse graining are retained.
    Fixed sampling parameter
    \(\varepsilon_{\mathrm{RG}}=2.8\) was used in this figure.
    } 
    \label{Fig. 3}
\end{figure*}

One advantage of ERRG is that it is applicable to diverse physical dynamics on networks. Here we demonstrate this using central-force mechanical networks, which feature vector potentials and direction-dependent edge constraints, and are beyond scalar graph Laplacian problems. 
This contrast highlights the central premise of ERRG: microscopic relevance of network components should be defined by the physical response, instead of only network topology.

We choose our examples to be mechanical networks with random holes  across different scales (Fig.~\ref{Fig. 3}a), which provide a controlled model of multiscale disordered materials~\cite{yang2017hierarchically}. 
For these two-dimensional (2D) vector networks, the edge projection is resolved along the axial and transverse directions of each edge, thus the edge Green's function $\mathcal{G}_e(s)$ is a \(2\times2\) matrix. We take the corresponding trace for $\chi_e$ to define our edge scores, 
as detailed in Supplementary Information Section~\ref{SI:edge_projection_vector}. This generalization changes the local response representation without changing the ERRG sampling and contraction procedure.

Again, ERRG accurately preserves the large-scale mechanical response, although the achievable node reduction is more limited than that in the scalar networks. 
The two-patch test and ensemble statistics show that the dominant vector response is preserved with an approximately \(10\%\) shape-response error at \(\eta_{\ell}\simeq0.44\) (Fig.~\ref{Fig. 3}b,c). 
The spectral diagnostics expose the origin of this limitation (Fig.~\ref{Fig. 3}d--f). Although the rescaled low-frequency sector remains in approximate agreement across RG levels, additional zero or near-zero modes appear during coarse graining and produce an elevated long-\(\widehat{\tau}\) heat-kernel entropy plateau. Thus, preservation of the prescribed vector response does not by itself guarantee preservation of mechanical rigidity. Stronger coarse graining of central-force networks may therefore require a rigidity-preserving constraint in addition to the response-based edge score.
Furthermore, as in the scalar networks, distributions all become increasingly  wider during ERRG (Fig.~\ref{Fig. 3}g-i), depicting the emergence of heterogeneity across ERRG. 
\subsection*{{Experimentally characterized nanomaterial networks}}

\begin{figure*}[t]
    \centering
    \includegraphics[width=.9\textwidth]{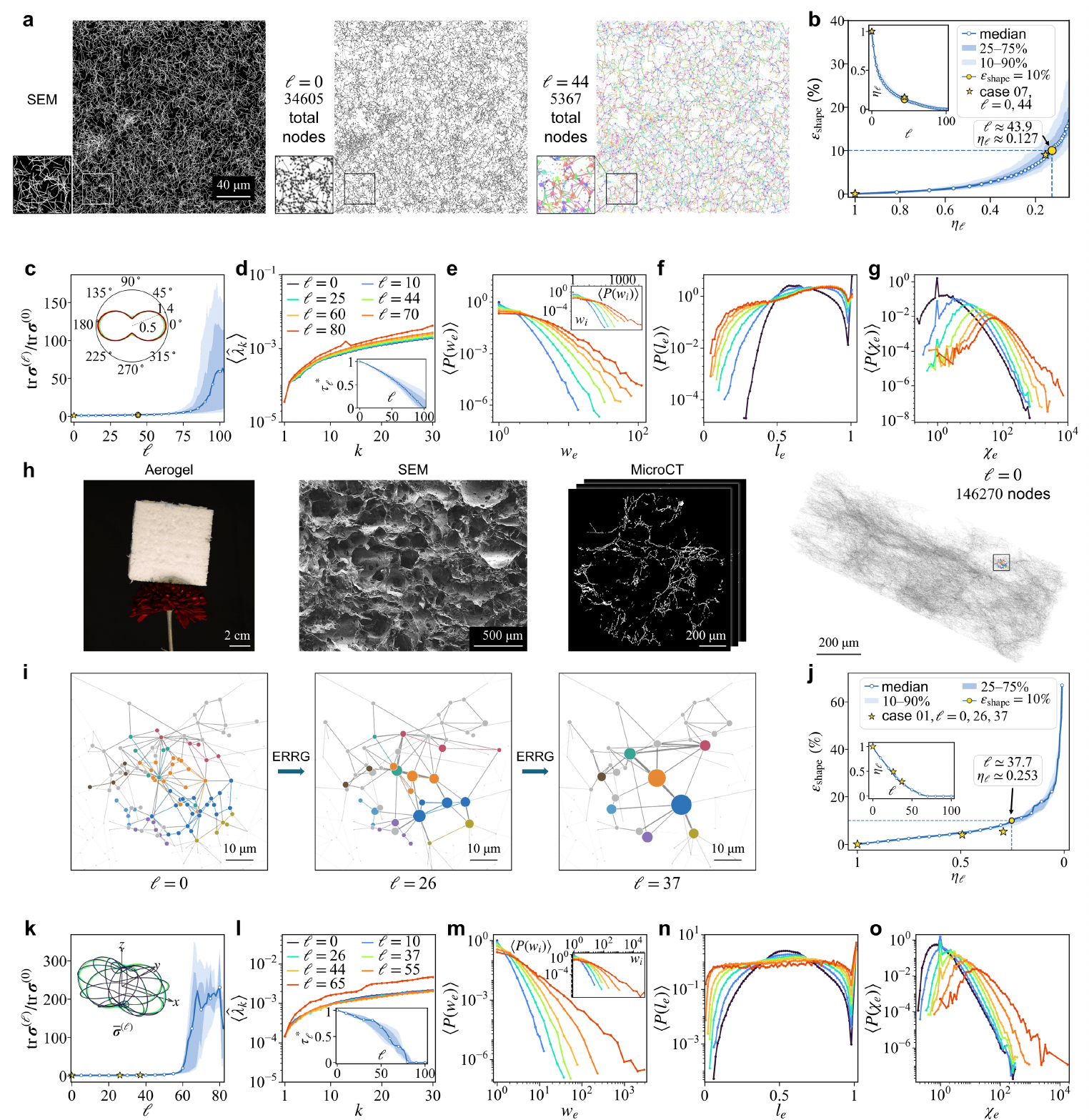}
    \caption{
    \textbf{ERRG flow of experimentally characterized nanomaterial networks.}
    \textbf{a--g,} SEM-derived nanowire networks.
    \textbf {a,} Representative SEM image and corresponding ERRG trajectory.
    Colors in the coarse network are categorical labels distinguishing different merged supernodes and are inherited across RG levels for visual tracking.
    \textbf{b,} Ensemble error--compression tradeoff over  %
    90 SEM-derived networks. 
    \textbf{c,} Evolution of the effective conductivity tensor \(\boldsymbol{\sigma}\): the ensemble median of \(\operatorname{tr}\boldsymbol{\sigma}^{(\ell)} /\operatorname{tr}\boldsymbol{\sigma}^{(0)}\) (main) and the trace-normalized directional effective conductivity at selected RG levels (inset).
    \textbf{d--g,} Ensemble-averaged network measures (median for \(\tau_\ell^*\)) across RG steps
    \(\ell\), including rescaled low-order eigenvalues and time-rescaling factor
    (d), edge and node weight distributions (e), edge leverage score
    distribution (f), and edge-weight regularized response time distribution
    (g).
    \textbf{h--o,} Experimental data and reconstructed networks from segmented microCT images for 3D ethyl cellulose aerogel networks.
    \textbf{h,} From left to right: photograph of the aerogel resting on a flower, cross-sectional SEM image, stacking representative microCT slices, and a skeletonized overview of the reconstructed network.
    \textbf{i,} magnified views of the highlighted region at RG levels 0, 26 and 37. Categorical colors track selected node groups through contraction. Scale bars, 2 cm (photograph), 500 \ensuremath{\mu}m (SEM), 200 \ensuremath{\mu}m (microCT slices and network overview), and 10 \ensuremath{\mu}m (magnified region).
    \textbf{j,} Ensemble error--compression tradeoff and node-count ratio (inset) for the five aerogel networks.
    \textbf{k,} Evolution of the effective conductivity tensor
    \(\boldsymbol{\sigma}\).
    \textbf{l--o,} Network measures across RG steps \(\ell\), including
    rescaled low-order eigenvalues and time-rescaling factor (l), edge and node
    weight distributions (m), edge leverage score distribution (n), and
    edge-weight regularized response time distribution (o).
    All ERRG trajectories use the fixed sampling parameter \(\varepsilon_{\mathrm{RG}}=3.8\).
    }
    \label{Fig. 4}
\end{figure*}

We next apply ERRG to two sets of technology-relevant complex materials. The first set consists of 2D nanowire networks obtained from scanning electron microscopy (SEM) (Fig.~\ref{Fig. 4}a). 
Fabricated through scalable layer-by-layer assembly, these networks form lightweight, flexible conductive coatings that can be used for electromagnetic shielding of robotic devices~\cite{wu2025layer}. 
The second set comprises five 3D ethyl cellulose aerogel networks reconstructed from segmented microCT images (Fig.~\ref{Fig. 4}h). 
These networks are made from polymer nanosheets formed by a scalable technique of antisolvent precipitation under shear, where polymer “nanoskins” peel off the solution droplets~\cite{bang2023fluid}. The nanosheets
then assemble into ultralow-density aerogels via foaming and freeze-drying  (see Supplementary Information Section~\ref{3D aerogel networks}).

These two experimentally assembled and characterized systems provide stringent tests of ERRG because their technologically important properties are intimately tied to their complex network architectures. In 2D nanowire networks, high electrical conductance at low surface coverage requires efficient conductive pathways, while in 3D nanosheet aerogels, mechanical stability at ultralow density relies on efficient stress pathways. More broadly, these materials exhibit complex hierarchical structures combining order and disorder, characterized by heterogeneous local coordination, anisotropy, pores, bottlenecks, and extended filamentary pathways. Together with their large sizes (typically \(10^4\)–\(10^5\) nodes), these structural features present a challenging test of whether ERRG can construct a multiscale effective representation directly from experimentally obtained microscopy images. 

We obtain the network structures from these experiments using an image-to-graph workflow, StructuralGT~\cite{vecchio2021structural,kadar2024structuralgt}. 
In the resulting graph, junctions and endpoints of the skeletonized nanowire or nanosheet aerogel networks define nodes, and the segments connecting them define edges. We assign uniform weights to all edges.

We then perform ERRG on these experimental networks using \(\varepsilon_{\mathrm{RG}}=3.8\) and scalar potential dynamics. 
Figure~\ref{Fig. 4}a,b,i,j show that ERRG substantially reduces the material  networks while retaining their physical response. 
As in the synthetic examples, the low-order spectra of these experimental-derived networks retain
their profiles after $\tau^*$ rescaling (Fig.~\ref{Fig. 4}d,l), while the edge
and node weight distributions and the distributions of \(l_e\)
and \(\chi_e\) evolve substantially during ERRG
(Fig.~\ref{Fig. 4}e--g,m--o), characterizing the renormalization flow across scales. 
One interesting feature of the 2D nanowire networks is their anisotropy: filaments exhibit a nematic order along the horizontal direction, a feature controllable through the preparation protocol. This shows up in the physical response as an anisotropic conductivity tensor $\boldsymbol{\sigma}$.  Although the overall conductivity $\textrm{tr} \boldsymbol{\sigma} $ exhibits an increase over ERRG iterations due to edge merging, the anisotropy is accurately preserved through ERRG (Fig.~\ref{Fig. 4}c). 
We show similar anisotropy characterization of the 3D ethyl cellulose aerogel network in Fig.~\ref{Fig. 4}k.

\begin{figure}[t]
    \centering
    \includegraphics[width=.9\columnwidth]{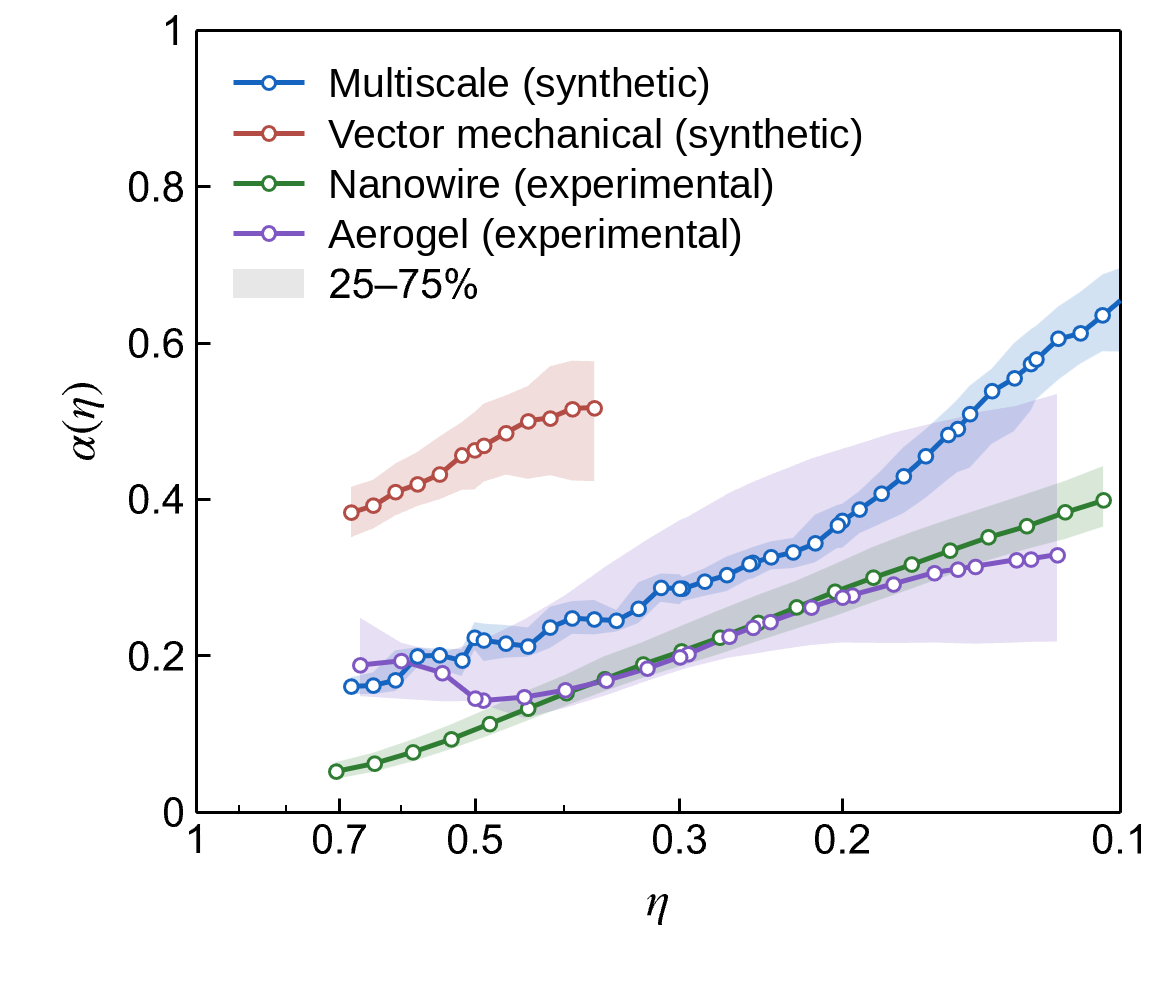}
    \caption{
    \textbf{Scale-dependent dynamical exponent $\alpha$ (Eq.~\eqref{EQ:alpha}) along the ERRG flow.}
    The analysis includes 90 synthetic multiscale physical networks
    (Fig.~\ref{Fig. 2}), 90 central-force mechanical networks (Fig.~\ref{Fig. 3}),
    90 nanowire networks (Fig.~\ref{Fig. 4}a), and 5 ethyl cellulose aerogel
    networks (Fig.~\ref{Fig. 4}h).
    For each sample, we take the median $\alpha(\eta)$ over 5 ERRG
    runs. Curves and shaded bands show the median and interquartile
    range, respectively, of these values across samples.
    }
    \label{fig:alpha_eta}
\end{figure}

\section*{Implications for multiscale physical networks}

We introduce ERRG, a
response-defined coarse-graining framework for heterogeneous physical
networks. 
ERRG uses edge-resolved Green's functions
to determine how microscopic degrees of freedom should be coarse-grained. 
Stochastic edge sampling converts this response
hierarchy into contraction blocks, while Kadanoff supernode projection produces an
iterable hierarchy of effective networks.
Across synthetic and experimentally characterized nanomaterial
networks, ERRG substantially reduces network size while preserving the response, providing information about property evolution across scales. Comparing with other network renormalization methods, we show that ERRG is both efficient in progressively compressing large networks while preserving physical response (see Supplementary Information Section~\ref{SI:nanowire_method_comparison}). 

The ERRG flow of the spectral timescale $\tau^*(\eta)$ and its associated scaling exponent 
\begin{equation}\label{EQ:alpha}
\alpha=\frac{d \ln \tau^*}{d \ln \eta}
\end{equation}
reveal the multiscale dynamical signatures of each network.  To interpret this scaling, we compare to dynamical field theory, where a dynamical exponent $z$ characterizes how characteristic time scales with length, 
$\tau\sim l^{z}$~\cite{Martin1973,Hohenberg1977}. 
As nodes merge during ERRG, $\eta=\eta_\ell=N_\ell/N_0$, we observe that the Laplacian eigenvalues $\widetilde\lambda$ shift up while \emph{maintaining the spectral shape with only an overall time rescaling $\tau^*(\eta)$}. This indicates that the underlying structure of the theory is maintained but with evolving exponents.  
While the exact relation between $\alpha$ and $z$ may not be immediately clear yet, we note that they are deeply connected and both describe the evolution of timescale across RG.  
Interestingly, we observe different values of scale dependence of the exponent $\alpha$ for different types of physical networks we study (Fig.~\ref{fig:alpha_eta}). These differing trends reveal distinct multiscale features of these networks, depicting how their coarse grained dynamics evolve across scales.  Although further analysis is  needed to fully reveal the physical meaning of this exponent, these initial results are showing its promise as a multiscale fingerprint for complex physical networks. 

It is  worth noting that ERRG operates within a quadratic description of the network, corresponding to linear response. The nontrivial scaling observed here therefore probes the multiscale organization of network structures rather than fluctuation corrections. 
It is also useful to discuss the relation between ERRG and real-space RG studies of criticality on diluted lattice models, such as percolation or rigidity percolation~\cite{Reynolds1977PercolationRG,Barre2009RigidityRG,StinchcombeThorpe2011RigidityRG}. Although disordered, these lattice percolation models are still supported by a homogeneous underlying space and thus the \emph{disorder is uncorrelated}. 
Therefore these models eventually feature a divergent length scale in the traditional sense of critical phenomena, which directs the critical exponents. The physics we discuss here instead focuses on \emph{correlated disorder}, where the connectivity of the network features multiscale complex organization, and the length scale needs a new definition.  The meaning of an RG flow in these complex networks is just beginning to be understood.

Furthermore, ERRG opens new possibilities to engineer complex physical networks utilizing renormalization as a powerful protocol, which is critical for attaining presently unachievable electrical, thermal, optical, and mechanical properties. 
Because ERRG is defined on a physically measurable response, it can be generalized to include nonlinear and nonreciprocal interactions, thermal fluctuations and active driving, as well as incorporating additional rules such as competing observables and constraints. 
A particularly promising direction is its application to biological and artificial neural networks, where ERRG may uncover low-dimensional effective structures that preserve learned functionality and provide new insights into the relationship between network architecture and learning dynamics. As a proof of concept, we demonstrate in Supplementary Information Section~\ref{sec:errg-task-preservation} that ERRG can substantially simplify a physical neural network while preserving its learned functionality with high accuracy. Exploring the richer structures and dynamics of more complex neural networks remains an exciting direction for future research.

\begin{acknowledgments}
We thank Mark Newman, William Stephenson, Chang Shu, and Yitong Zhang for helpful discussions.
This work was supported in part by the National Science Foundation through the Center for Complex Particle Systems (COMPASS) under Award No.~2243104.
\end{acknowledgments}

\noindent\textbf{Author contributions.}

H.Y. and X.M. developed the theoretical framework. H.Y. carried out the theoretical and numerical analyses, 
and drafted the manuscript. L.S. and W.W. prepared the nanowire networks and performed their graph extraction. S.L. prepared the physical neural networks. B.K. and W.W. prepared ethyl cellulose aerogels and performed their graph extraction. N.K. and O.D.V. contributed to the scientific insight and interpretation of results. 
All authors contributed to the discussions and writing of the paper. X.M. supervised the work. 

\bibliographystyle{unsrt}
\bibliography{references}

\appendix
\subsection{Edge projection for physical networks}
\label{SI:edge_projection_vector}

For an oriented edge \(e=(i,j)\), the edge projection vector maps nodal potentials to an edge potential drop,
\[
\varepsilon_e=\mathbf b_e^\top\mathbf u .
\]
Changing the edge orientation changes \(\mathbf b_e\to-\mathbf b_e\), but does not affect self-response quantities such as \(\mathbf b_e^\top\mathbf G(s)\mathbf b_e\).

For a scalar network, \(\mathbf u=(u_1,\ldots,u_N)^\top\). 
The vector \(\mathbf b_e\) has entry \(-1\) on node \(i\), entry \(+1\) on node \(j\), and zero elsewhere, giving
\[
\varepsilon_e
=
\mathbf b_e^\top\mathbf u
=
u_j-u_i .
\]
An edge-dipolar load of amplitude \(p(t)\) is written as \(\mathbf b_e p(t)\).

For a two-dimensional vector network, the relative displacement across an edge has two independent local components. 
We choose an orthonormal basis formed by the axial direction along the bond and the transverse direction perpendicular to the bond. 
Let \(\hat{\mathbf n}_{ij}=(n_x,n_y)\) be the unit vector from node \(i\) to node \(j\), and let \(\hat{\mathbf t}_{ij}=(-n_y,n_x)\) be a transverse unit vector. 
The axial projection vector has block \(-\hat{\mathbf n}_{ij}\) on node \(i\), block \(+\hat{\mathbf n}_{ij}\) on node \(j\), and zero blocks elsewhere. 
Equivalently,
\[
\mathbf b_{e,\parallel}^{\top}
=
(0,\ldots,-n_x,-n_y,\ldots,+n_x,+n_y,\ldots,0).
\]
The transverse projection vector is defined in the same way using \(\hat{\mathbf t}_{ij}\),
\[
\mathbf b_{e,\perp}^{\top}
=
(0,\ldots,+n_y,-n_x,\ldots,-n_y,+n_x,\ldots,0).
\]
These give the axial and transverse edge responses
\[
\varepsilon_{e,\parallel}
=
\mathbf b_{e,\parallel}^{\top}\mathbf u
=
\hat{\mathbf n}_{ij}\cdot(\mathbf u_j-\mathbf u_i),
\qquad
\varepsilon_{e,\perp}
=
\mathbf b_{e,\perp}^{\top}\mathbf u
=
\hat{\mathbf t}_{ij}\cdot(\mathbf u_j-\mathbf u_i).
\]
This follows from the mechanical compatibility matrix construction described in Ref.~\cite{pellegrino1986matrix,mao2018maxwell} and can be straighforwardly extended to three-dimensional networks. 

The corresponding axial and transverse edge Green's functions are
\[
\mathcal G_{e,\parallel}(s)
=
\mathbf b_{e,\parallel}^{\top}\mathbf G(s)\mathbf b_{e,\parallel},
\qquad
\mathcal G_{e,\perp}(s)
=
\mathbf b_{e,\perp}^{\top}\mathbf G(s)\mathbf b_{e,\perp}.
\]
More generally, these two local directions define a \(2\times2\) edge-local response matrix,
\[
\mathcal G_e^{\mathrm{loc}}(s)
=
\begin{pmatrix}
\mathbf b_{e,\parallel}^{\top}\mathbf G(s)\mathbf b_{e,\parallel}
&
\mathbf b_{e,\parallel}^{\top}\mathbf G(s)\mathbf b_{e,\perp}
\\
\mathbf b_{e,\perp}^{\top}\mathbf G(s)\mathbf b_{e,\parallel}
&
\mathbf b_{e,\perp}^{\top}\mathbf G(s)\mathbf b_{e,\perp}
\end{pmatrix}.
\]
The matrix \(\mathcal G_e^{\mathrm{loc}}(s)\) is the vector-edge Green's function referred to in the main text. The scalar response used for trace scoring is its trace,
\[
\mathcal G_{e,\mathrm{tr}}(s)
=
\operatorname{tr}\mathcal G_e^{\mathrm{loc}}(s)
=
\mathcal G_{e,\parallel}(s)
+
\mathcal G_{e,\perp}(s).
\]
Thus, although one can probe infinitely many directions, in two dimensions all such probes are linear combinations of the axial and transverse channels. 
The trace response is obtained by summing the self-responses over these two orthogonal directions. 
In \(d\) dimensions, the same construction generalizes by taking the trace over one axial direction and \(d-1\) transverse directions.

\subsection{Shifted steady-state measurements of response moments}
\label{SI:shifted_steady_response}

Here we explain why the spatiotemporal edge response used in the main text can be obtained from shifted steady-state measurements. We first consider the unit-drag special case \(\boldsymbol{\Gamma}=\mathbf I\), for which \(\mathbf{\hat A}(s)=s\mathbf I\). In this case the shifted operator is \(\mathbf L+\alpha_{\mathrm{leak}}\mathbf I\), with \(\alpha_{\mathrm{leak}}\ge 0\). The parameter \(\alpha_{\mathrm{leak}}\) represents a uniform leakage rate and acts as a controllable time-scale filter.

In an electrical network, adding \(\alpha_{\mathrm{leak}}\mathbf I\) corresponds to connecting each node to ground through the same shunt conductance ~\cite{gurfinkel2021adjustable}. Changing the grounding conductance changes \(\alpha_{\mathrm{leak}}\), so the leakage dependence of the measured edge voltage drop gives \(\tau_e\) through the finite-difference procedure below. Analogous implementations apply whenever the shifted term can be realized as a uniform local leakage or pinning, such as leakage to the environment in diffusion or thermal networks, or weak uniform pinning of displacement degrees of freedom to a fixed background in vector mechanical networks.

Consider applying a constant edge-dipolar load on edge \(e\). 
In the presence of uniform leakage \(\alpha_{\mathrm{leak}}\), the steady-state response is governed by
\[
(\mathbf L+\alpha_{\mathrm{leak}}\mathbf I)\mathbf u=\mathbf b_e .
\]
The edge readout is therefore
\[
\mathcal G_e(\alpha_{\mathrm{leak}})
=
\mathbf b_e^{\top}\mathbf u
=
\mathbf b_e^{\top}
(\mathbf L+\alpha_{\mathrm{leak}}\mathbf I)^{-1}
\mathbf b_e .
\]
Thus the shifted Green's function \(\mathcal G_e(\alpha_{\mathrm{leak}})\) can be measured directly as a steady-state response, without recording the full time-dependent response field.

This steady-state measurement is equivalent to an exponentially filtered time integral of the unshifted response kernel. 
For \(\alpha_{\mathrm{leak}}>0\),
\[
(\mathbf L+\alpha_{\mathrm{leak}}\mathbf I)^{-1}
=
\int_0^\infty e^{-t(\mathbf L+\alpha_{\mathrm{leak}}\mathbf I)}\,dt ,
\]
and therefore
\[
\mathcal G_e(\alpha_{\mathrm{leak}})
=
\int_0^\infty
e^{-\alpha_{\mathrm{leak}} t}
\mathbf b_e^{\top}
e^{-t\mathbf L}
\mathbf b_e
\,dt .
\]
Small \(\alpha_{\mathrm{leak}}\) retains slow, long-time relaxation modes, whereas large \(\alpha_{\mathrm{leak}}\) emphasizes short-time, more local response components. 
In the zero-leakage limit, understood on the gauge-fixed subspace,
\[
\mathcal G_e(0)
=
\mathbf b_e^{\top}\mathbf L^+\mathbf b_e
=
R_e ,
\]
which is the effective resistance, or effective compliance in the mechanical setting.

\begin{figure*}[t]
    \centering
    \includegraphics[width=\textwidth]{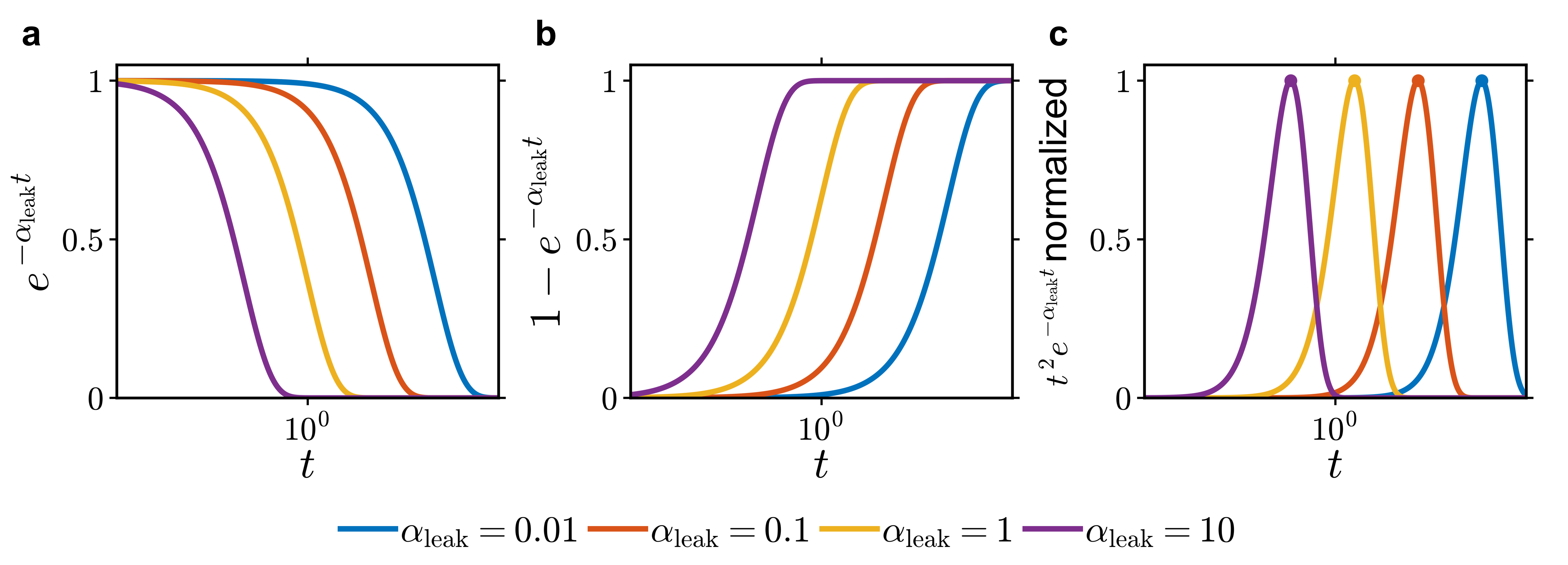}
    \caption{
    \textbf{Temporal filters for shifted steady-state measurements.}
    \textbf{a,} Exponential cutoff \(e^{-\alpha_{\mathrm{leak}} t}\).
    \textbf{b,} Accumulation factor \(1-e^{-\alpha_{\mathrm{leak}} t}\).
    \textbf{c,} Peak-normalized temporal window \(t^2e^{-\alpha_{\mathrm{leak}} t}\).
    }
    \label{fig:temporal_filters}
\end{figure*}

Response moments follow from derivatives of the shifted response with respect to \(\alpha_{\mathrm{leak}}\). 
For integer \(n\ge 0\),
\[
\mathcal M_e^{(n)}(\alpha_{\mathrm{leak}})
=
\frac{1}{n!}
\int_0^\infty
t^n e^{-\alpha_{\mathrm{leak}} t}
\mathbf b_e^{\top}
e^{-t\mathbf L}
\mathbf b_e
\,dt ,
\]
or equivalently,
\[
\mathcal M_e^{(n)}(\alpha_{\mathrm{leak}})
=
\mathbf b_e^{\top}
(\mathbf L+\alpha_{\mathrm{leak}}\mathbf I)^{-(n+1)}
\mathbf b_e
=
\frac{(-1)^n}{n!}
\frac{\partial^n}{\partial \alpha_{\mathrm{leak}}^n}
\mathcal G_e(\alpha_{\mathrm{leak}}).
\]
The factor \(t^n e^{-\alpha_{\mathrm{leak}} t}\) defines a tunable temporal filter. The exponential factor suppresses late-time response, while the polynomial factor shifts the dominant contribution toward a finite response-time window, with characteristic scale \(t\sim n/\alpha_{\mathrm{leak}}\) for \(n>0\). Thus increasing \(\alpha_{\mathrm{leak}}\) emphasizes shorter-time response components, whereas increasing \(n\) probes later response windows. This is illustrated in Fig.~\ref{fig:temporal_filters}, where panel (a) shows the exponential cutoff \(e^{-\alpha_{\mathrm{leak}} t}\), panel (b) shows the accumulation factor \(1-e^{-\alpha_{\mathrm{leak}} t}\), and panel (c) shows the peak-normalized finite-window factor \(t^2 e^{-\alpha_{\mathrm{leak}} t}\). The accumulation factor \(1-e^{-\alpha_{\mathrm{leak}} t}\) illustrates approach to a steady-state value over \(t\sim1/\alpha_{\mathrm{leak}}\), while the \(n=2\) window peaks at \(t=2/\alpha_{\mathrm{leak}}\).

In particular,
\[
\mathcal M_e^{(0)}(0)=R_e ,
\qquad
\mathcal M_e^{(1)}(0)
=
\mathbf b_e^{\top}
(\mathbf L^+)^2
\mathbf b_e .
\]
In this unit-drag special case, the response time used in the main text can therefore be written as 
\[
\tau_e
=
\frac{\mathcal M_e^{(1)}(0)}
{\mathcal M_e^{(0)}(0)}
=
-\left.
\frac{\partial}{\partial \alpha_{\mathrm{leak}}}
\log \mathcal G_e(\alpha_{\mathrm{leak}})
\right|_{\alpha_{\mathrm{leak}}=0}.
\]
Estimating \(\tau_e\) requires only two steady-state measurements per edge, one at zero leakage and one at weak leakage.
For small positive \(\Delta\alpha_{\mathrm{leak}}\),
\[
\tau_e
\simeq
\frac{
\log \mathcal G_e(0)-\log \mathcal G_e(\Delta\alpha_{\mathrm{leak}})
}{
\Delta\alpha_{\mathrm{leak}}
}.
\]

For general nodal drag, we use the definition in the main text,
\[
\tau_e=
\frac{\mathbf b_e^{\top}\mathbf L^+\boldsymbol{\Gamma}
\mathbf L^+\mathbf b_e}
{\mathbf b_e^{\top}\mathbf L^+\mathbf b_e}.
\]
The unit-drag result above is recovered by setting
\(\boldsymbol{\Gamma}=\mathbf I\). Accumulated coarse-node drags
are retained in \(\boldsymbol{\Gamma}^{(\ell)}\); they are not
assumed to remain equal after contraction.

\subsection{Response-biased edge sampling and contraction}
\label{SI:response_biased_sampling}

Here we describe the randomized sampling procedure used to implement one ERRG step. 
At a given RG level, each edge \(e\in N_e\) is assigned a response score \(E_e\). 
Edges with large \(E_e\) represent heavily weighted pathways for slow responses and should therefore be protected during coarse graining, whereas edges with small \(E_e\) are more contraction-prone. 
ERRG implements this separation using a randomized edge-sampling procedure inspired by spectral sparsification \cite{spielman2008graph}.

Given the edge scores, we define the normalized sampling probability
\[
p_e
=
\frac{E_e}
{\sum_{e'\in N_e}E_{e'}} .
\]
For a sampling budget \(q\), we perform \(q\) independent draws over the full edge set, with replacement, according to \(p_e\). 
Let \(X_e\) denote the number of times edge \(e\) is selected. 
For each edge, the marginal distribution is
\[
X_e
\sim
\mathrm{Binomial}\!\left(q,p_e\right).
\]
Edges selected at least once are retained,
\[
N_e^{\mathrm{keep}}
=
\{e\in N_e:X_e>0\},
\]
whereas edges that are not selected are placed in
\[
N_e^{\mathrm{omit}}
=
N_e\setminus N_e^{\mathrm{keep}} .
\]
In ERRG, these unsampled edges are used as contraction signals. 
The contraction blocks are defined as the connected components of the graph \((V,N_e^{\mathrm{omit}})\). 
Each connected component is merged into one coarse node, while edges between different components define the inter-block connections of the coarse graph. 
The weights of inter-block edges are obtained by summing the weights of the original edges connecting the corresponding blocks.

This sampling rule explains the exponential factor used in the response-biased edge ensemble. 
The probability that edge \(e\) is not selected in any of the \(q\) draws is
\[
\Pr[X_e=0]
=
(1-p_e)^q .
\]
For small \(p_e\), this gives
\[
\Pr[X_e=0]
\approx
\exp[-q\,p_e] .
\]
Using the normalized probability \(p_e=E_e/\sum_{e'}E_{e'}\), we obtain
\[
\Pr[X_e=0]
\approx
\exp(-\beta E_e),
\qquad
\beta
=
\frac{q}
{\sum_{e'\in N_e}E_{e'}} .
\]
Thus, the factor \(\exp(-\beta E_e)\) is not the direct sampling probability. 
It emerges as the approximate probability that an edge is omitted by the normalized sampling procedure. 
Consequently, low-\(E_e\) edges are more likely to be omitted and contracted, whereas high-\(E_e\) edges are more likely to be sampled and retained.

The sampling budget \(q\) sets the practical coarse-graining resolution. 
In the numerical implementation, we use a sparsification-style scaling \cite{spielman2008graph}
\[
q=q(\varepsilon_{\mathrm{RG}})
\sim
\frac{N\log N}{\varepsilon_{\mathrm{RG}}^2},
\]
where \(N\) is the number of nodes at the current RG level and \(\varepsilon_{\mathrm{RG}}\) is the ERRG sampling parameter. 
Smaller \(\varepsilon_{\mathrm{RG}}\) increases the number of sampled edges and therefore gives a finer coarse-graining step, whereas larger \(\varepsilon_{\mathrm{RG}}\) produces stronger compression.

The computation of the response scores is separate from the edge-sampling step. 
For the leverage score $l_e=w_eR_e$, the effective-resistance
factor $R_e$, and more generally the shifted response quantities
introduced in Sec.~\ref{SI:shifted_steady_response}, the required quantities are edge-wise quadratic forms of Green's operators. 
For large networks, these quadratic forms can be estimated efficiently using Laplacian solvers and Johnson--Lindenstrauss random projections, following standard nearly-linear-time effective-resistance estimation schemes \cite{spielman2004nearly,spielman2008graph}.
In the shifted setting, the same idea is applied to operators of the form \(\mathbf L+\alpha_{\mathrm{leak}}\mathbf I\), so that response quantities at different leakage values can be estimated from a small number of shifted linear solves rather than from one solve per edge. 
The solves for different random probes and different leakage parameters \(\alpha_{\mathrm{leak}}\) are independent and are therefore naturally parallelizable. 
This provides a scalable route for computing the response scores used by ERRG on large networks.

\subsection{Response error metrics}
\label{SI:response_error}

To compare responses across RG levels, the coarse response is first lifted back to the original node space. 
We write the fine response on the original graph as
\(\mathbf U^{(0)}\in\mathbb{R}^{N_0\times d}\), and the lifted response from RG level \(\ell\) as
\(\widetilde{\mathbf U}^{(\ell)}\in\mathbb{R}^{N_0\times d}\). 
Here \(d=1\) for scalar networks and \(d=2\) for vector mechanical networks. 
Equivalently, the vector response may be viewed as a flattened vector in \(\mathbb{R}^{dN_0}\).

Because the Laplacian or stiffness response is defined up to gauge modes, we first remove the reference component before comparison. 
For scalar networks this is the additive constant mode. 
For vector networks this corresponds to removing the uniform translation mode component-wise,
\[
\mathbf U^{(0)}_0
=
\mathbf U^{(0)}
-
\mathbf 1 \left\langle \mathbf U^{(0)}\right\rangle,
\qquad
\widetilde{\mathbf U}^{(\ell)}_0
=
\widetilde{\mathbf U}^{(\ell)}
-
\mathbf 1 \left\langle \widetilde{\mathbf U}^{(\ell)}\right\rangle ,
\]
where \(\langle \mathbf U\rangle\in\mathbb{R}^{1\times d}\) denotes the average response over nodes. 
If additional rigid-body modes are present in a mechanical network, the same gauge-fixing or null-space projection used in the response solve is applied before evaluating the following metrics.

The primary response error is the shape-response error,
\[
\varepsilon_{\mathrm{shape}}^{(\ell)}
=
\frac{
\left\|
\widetilde{\mathbf U}^{(\ell)}_0
-
\mathbf U^{(0)}_0
\right\|_F
}{
\left\|
\mathbf U^{(0)}_0
\right\|_F
}.
\]
For scalar networks, the Frobenius norm reduces to the usual vector \(L^2\) norm. 
For vector networks, it is computed over all nodal displacement components, equivalently after flattening the displacement field into a vector in \(\mathbb{R}^{dN_0}\). 
In the figures, this quantity is reported as a percentage, \(100\varepsilon_{\mathrm{shape}}\).

We also report the shape-cosine mismatch,
\[
1-\cos\theta_{\mathrm{shape}}^{(\ell)}
=
1-
\frac{
\left\langle
\mathbf U^{(0)}_0,
\widetilde{\mathbf U}^{(\ell)}_0
\right\rangle_F
}{
\left\|\mathbf U^{(0)}_0\right\|_F
\left\|\widetilde{\mathbf U}^{(\ell)}_0\right\|_F
},
\]
where
\[
\left\langle
\mathbf A,\mathbf B
\right\rangle_F
=
\sum_{i=1}^{N_0}
\mathbf A_i\cdot \mathbf B_i
\]
is the Frobenius inner product. 
For scalar response fields, \(\cos\theta_{\mathrm{shape}}^{(\ell)}\) is exactly the Pearson correlation coefficient between the mean-subtracted fine response and the lifted coarse response. 
Thus smaller \(1-\cos\theta_{\mathrm{shape}}^{(\ell)}\) indicates stronger preservation of the scalar response pattern. 
For vector mechanical networks, the same expression compares the flattened, gauge-fixed displacement fields through their normalized Frobenius inner product.

\subsection{Spectral and heat-kernel diagnostics}
\label{SI:spectral_heat_kernel}

For each RG level \(\ell\), we compute both bare-operator and
drag-normalized spectral diagnostics. The bare coarse operator is denoted by \(\mathbf L^{(\ell)}\). After removing the prescribed gauge zero mode, its remaining eigenvalues are written as
\[
\{\lambda_k^{(\ell)}\}_{k\ge1},
\]
where \(k=1\) labels the first nonzero mode. The corresponding normalized spectral density is denoted by
\[
D^{(\ell)}(\lambda).
\]

The bare-operator heat-kernel partition function and normalized spectral weights are
\[
Z_{\mathrm{HK}}^{(\ell)}(\tau)
=
\sum_{k\ge1}
e^{-\tau\lambda_k^{(\ell)}},
\]
and
\[
p_{k,\mathrm{HK}}^{(\ell)}(\tau)
=
\frac{
e^{-\tau\lambda_k^{(\ell)}}
}{
Z_{\mathrm{HK}}^{(\ell)}(\tau)
}.
\]
The corresponding heat-kernel entropy is
\[
S_{\mathrm{HK}}^{(\ell)}(\tau)
=
-\sum_{k\ge1}
p_{k,\mathrm{HK}}^{(\ell)}(\tau)
\log p_{k,\mathrm{HK}}^{(\ell)}(\tau),
\]
and the heat-kernel specific heat is
\[
C_{\mathrm{HK}}^{(\ell)}(\tau)
=
-\tau
\frac{dS_{\mathrm{HK}}^{(\ell)}(\tau)}{d\tau}
=
\tau^2
\operatorname{Var}_{\tau}^{(\ell)}(\lambda).
\]
These bare-operator quantities are used in the supplementary diagnostics to show the uncorrected spectral flow generated by ERRG.

Because each coarse node represents a block of fine nodes, the coarse overdamped dynamics also carries an effective drag matrix. Let
\(\mathbf Q^{(\ell)}\) denote the lifting matrix from RG level \(\ell\) to
the original node space. For unit drag on the fine nodes,
\[
\boldsymbol{\Gamma}^{(\ell)}
=
\left(\mathbf Q^{(\ell)}\right)^{\top}
\mathbf Q^{(\ell)}.
\]
The diagonal entries of \(\boldsymbol{\Gamma}^{(\ell)}\) give the numbers of fine nodes represented by the corresponding coarse nodes.

To account for this accumulated drag, we define the drag-normalized operator
\[
\widetilde{\mathbf L}^{(\ell)}
=
\left(\boldsymbol{\Gamma}^{(\ell)}\right)^{-1/2}
\mathbf L^{(\ell)}
\left(\boldsymbol{\Gamma}^{(\ell)}\right)^{-1/2}.
\]
Equivalently, its eigenvalues are obtained from the generalized eigenvalue problem
\[
\mathbf L^{(\ell)}\boldsymbol{\phi}_k
=
\widetilde{\lambda}_k^{(\ell)}
\boldsymbol{\Gamma}^{(\ell)}
\boldsymbol{\phi}_k.
\]
After removing the prescribed gauge mode, the drag-normalized relaxation rates are denoted by
\[
\{\widetilde{\lambda}_k^{(\ell)}\}_{k\ge1},
\]
with \(\widetilde{\lambda}_1^{(\ell)}\) the first nonzero drag-normalized eigenvalue. The corresponding normalized spectral density is written as
\[
D^{(\ell)}(\widetilde{\lambda}).
\]

The drag-normalized operator is used only for spectral and dynamical diagnostics. The static response comparisons use the physical coarse operator \(\mathbf L^{(\ell)}\), together with the corresponding coarse loading and lifting map.

To compare the characteristic relaxation scales across RG levels, we define
\[
\tau_\ell^*
=
\frac{
\widetilde{\lambda}_1^{(0)}
}{
\widetilde{\lambda}_1^{(\ell)}
},
\]
and introduce the rescaled eigenvalues and heat-kernel time
\[
\widehat{\lambda}_k^{(\ell)}
=
\tau_\ell^*
\widetilde{\lambda}_k^{(\ell)},
\qquad
\widehat{\tau}
=
\frac{\tau}{\tau_\ell^*}.
\]
These definitions preserve the heat-kernel argument,
\[
\widehat{\tau}\,
\widehat{\lambda}_k^{(\ell)}
=
\tau\,
\widetilde{\lambda}_k^{(\ell)}.
\]

Let \(D^{(\ell)}(\widehat{\lambda})\) denote the normalized density of the rescaled eigenvalues. To account for the reduction in the number of retained spectral modes, the mode-count-corrected density shown in the main text is defined as
\[
\overline{D}^{(\ell)}(\widehat{\lambda})
=
\frac{N_+^{(\ell)}}{N_+^{(0)}}
D^{(\ell)}(\widehat{\lambda}),
\]
where \(N_+^{(\ell)}\) is the number of nonzero modes included at RG level \(\ell\). For a connected scalar network, \(N_+^{(\ell)}=N_\ell-1\).

Using the rescaled spectrum, the heat-kernel weights shown in the main-text
figures are defined directly as
\[
p_{k,\mathrm{HK}}^{(\ell)}(\widehat{\tau})
=
\frac{
e^{-\widehat{\tau}\widehat{\lambda}_k^{(\ell)}}
}{
\sum_j e^{-\widehat{\tau}\widehat{\lambda}_j^{(\ell)}}
}.
\]
The corresponding heat-kernel entropy and specific heat are
\[
S_{\mathrm{HK}}^{(\ell)}(\widehat{\tau})
=
-\sum_k
p_{k,\mathrm{HK}}^{(\ell)}(\widehat{\tau})
\log p_{k,\mathrm{HK}}^{(\ell)}(\widehat{\tau}),
\]
and
\[
C_{\mathrm{HK}}^{(\ell)}(\widehat{\tau})
=
-\widehat{\tau}
\frac{
dS_{\mathrm{HK}}^{(\ell)}(\widehat{\tau})
}{
d\widehat{\tau}
}
=
\widehat{\tau}^{\,2}
\operatorname{Var}_{\widehat{\tau}}^{(\ell)}
(\widehat{\lambda}).
\]
Because
\[
\widehat{\tau}\widehat{\lambda}_k^{(\ell)}
=
\tau\widetilde{\lambda}_k^{(\ell)},
\]
the rescaling changes the characteristic time axis without changing the
heat-kernel exponential weights.

For vector central-force networks, the same definitions are applied with \(\mathbf L^{(\ell)}\) replaced by the stiffness operator.

\subsection{Effective conductivity tensor}
\label{SI:effective_tensor}

To quantify the macroscopic transport response of the SEM-derived
networks, we compute the two-dimensional effective conductivity tensor
at each RG level \(\ell\),
\[
\boldsymbol{\sigma}^{(\ell)}
=
\begin{bmatrix}
\sigma_{xx}^{(\ell)} & \sigma_{xy}^{(\ell)}\\
\sigma_{xy}^{(\ell)} & \sigma_{yy}^{(\ell)}
\end{bmatrix}.
\]

The same physical SEM window and the same set of fine-network boundary
nodes are used to define the boundary conditions at every RG level.
Two affine Dirichlet problems are solved on each coarse network,
corresponding to unit potential gradients in the \(x\) and \(y\)
directions. Let \(\mathbf u_x^{(\ell)}\) and \(\mathbf u_y^{(\ell)}\)
denote the resulting potential fields. The effective conductivity
tensor is calculated from the bilinear dissipation per unit area of
the fixed SEM window,
\[
\boldsymbol{\sigma}^{(\ell)}
=
\frac{1}{A_{\mathrm{SEM}}}
\begin{bmatrix}
\left(\mathbf u_x^{(\ell)}\right)^T
\mathbf L^{(\ell)}\mathbf u_x^{(\ell)}
&
\left(\mathbf u_x^{(\ell)}\right)^T
\mathbf L^{(\ell)}\mathbf u_y^{(\ell)}
\\
\left(\mathbf u_y^{(\ell)}\right)^T
\mathbf L^{(\ell)}\mathbf u_x^{(\ell)}
&
\left(\mathbf u_y^{(\ell)}\right)^T
\mathbf L^{(\ell)}\mathbf u_y^{(\ell)}
\end{bmatrix},
\]
where \(A_{\mathrm{SEM}}\) is the area of the fixed SEM window.
The resulting tensor is symmetric up to numerical roundoff.

The 2D main panel reports
\[
\frac{\operatorname{tr}\boldsymbol{\sigma}^{(\ell)}}
     {\operatorname{tr}\boldsymbol{\sigma}^{(0)}},
\]
which measures the change in the overall magnitude of the effective
conductivity relative to the fine network.

The 2D inset shows the directional dependence of the effective
conductivity. At polar angle \(\theta\), the radial coordinate is
\[
\frac{
\sigma_{xx}^{(\ell)}\cos^2\theta
+
2\sigma_{xy}^{(\ell)}\sin\theta\cos\theta
+
\sigma_{yy}^{(\ell)}\sin^2\theta
}{
\operatorname{tr}\boldsymbol{\sigma}^{(\ell)}/2
}.
\]
The numerator is the effective conductivity along the direction
specified by \(\theta\), whereas the denominator is the mean of the two
principal effective conductivities at the same RG level. This
normalization removes the overall conductivity magnitude and isolates
the directional anisotropy of the effective response. An isotropic
effective tensor produces a circular polar curve of unit radius. An
elongated curve indicates anisotropic transport, with the long axis
corresponding to the direction of larger effective conductivity.

For the three-dimensional networks reconstructed from microCT images,
we extend the affine Dirichlet construction to the \(x\), \(y\), and \(z\)
directions. For each network, the boundary nodes are selected once at
\(\ell=0\) using the axis-aligned bounding box of the fine network.
For each of the six faces, we select all fine nodes within a distance
\(0.04L_i\) of the face, where \(L_i\) is the box length along the axis
normal to that face. If fewer than four nodes lie in this band, we
select the four closest nodes to the face instead. The union of these
six sets defines the fine-network boundary.

At each RG level, a coarse node is assigned a Dirichlet condition if it
contains at least one fine-network boundary node. For the unit-gradient
problem in direction \(i\), the prescribed potential at an original
boundary node \(p\) is \(-\bigl(r_{p,i}^{(0)}-\bar r_i^{(0)}\bigr)\),
where \(r_{p,i}^{(0)}\) is its coordinate along \(i\) and
\(\bar r_i^{(0)}\) is the mean coordinate of all fine nodes along that
axis. If several original boundary nodes map to the same coarse node,
their prescribed potentials are averaged. Interior members of that
coarse node do not enter the average.

The components of the resulting \(3\times3\) tensor are
\[
\sigma_{ij}^{(\ell)}
=
\frac{
\left(\mathbf u_i^{(\ell)}\right)^T
\mathbf L^{(\ell)}\mathbf u_j^{(\ell)}
}{V_0},
\qquad i,j\in\{x,y,z\},
\]
where \(V_0=L_xL_yL_z\) is the fine-network bounding-box volume, held
fixed across RG levels. The 3D main panel reports the median of the
trace ratios across the five networks, each normalized by its own
fine-level trace.

The 3D inset shows the directional conductivity surface for case 01
at \(\ell=0,26,37\). Its radius along a unit direction \(\hat{\mathbf n}\)
is \(\hat{\mathbf n}^{\top}\bar{\boldsymbol{\sigma}}^{(\ell)}\hat{\mathbf n}\),
where the trace-normalized conductivity tensor is
\[
\bar{\boldsymbol{\sigma}}^{(\ell)}
= \frac{\boldsymbol{\sigma}^{(\ell)}}
       {\operatorname{tr}\boldsymbol{\sigma}^{(\ell)}}.
\]
With this normalization, an isotropic tensor gives a sphere of radius
\(1/3\). Thus, in both dimensions, the main panel tracks the overall
conductivity magnitude, while the inset isolates directional anisotropy.

\subsection{Multiscale physical network ensemble and ERRG diagnostics}
\label{SI:voronoi_ensemble}

We generate 90 independent multiscale physical network realizations using the same construction protocol with different random seeds. For each realization, 70 Voronoi sites are randomly placed in a square domain, and the finite Voronoi ridges generated by these sites define the large-scale skeleton. The skeleton is broadened into a tubular region with a reference radius of \(1.2\), including weak segment-to-segment width variations and smooth thickening near multi-branch junctions. Exactly \(N_0=4096\) nodes are sampled within this region and relaxed to reduce local clustering. Candidate connections are constructed by Delaunay triangulation and filtered according to their length and whether their midpoint lies within the prescribed region. Any disconnected components remaining after filtering are joined through their shortest intercomponent links. The coordinates are finally rescaled to a common mean edge length, which does not alter the graph topology. All edges have uniform weight, so variations in the response and ERRG edge scores arise solely from network geometry and connectivity.

\begin{figure*}[t]
    \centering
    \includegraphics[width=\textwidth]{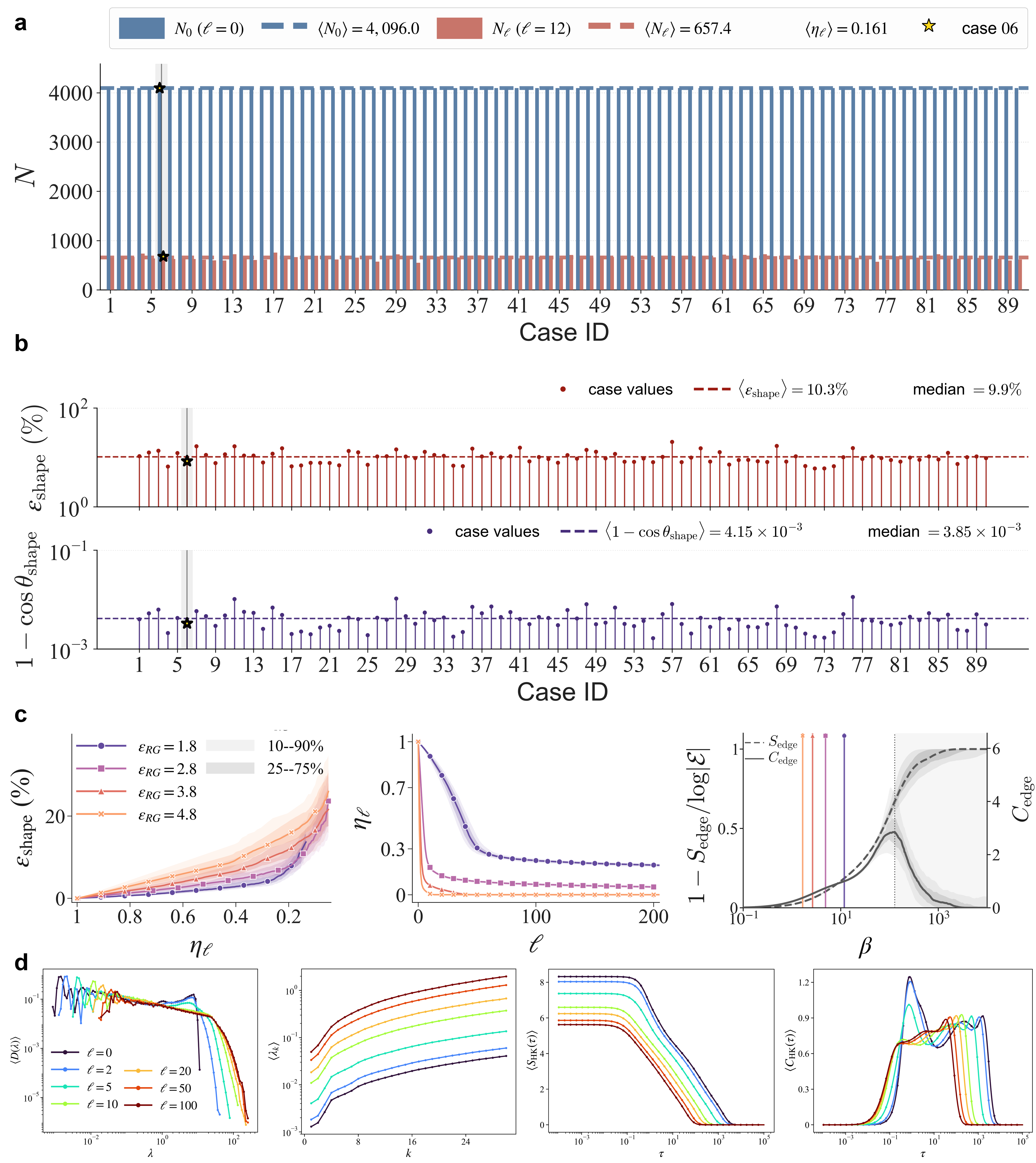}
    \caption{
    \textbf{Ensemble robustness, parameter selection, and bare-Laplacian
    diagnostics for multiscale physical networks.}
    \textbf{a,b,} Case-wise compression and scalar-response metrics for
    90 realizations at \(\ell=12\), using
    \(\varepsilon_{\mathrm{RG}}=2.8\). Yellow stars identify case 06.
    \textbf{c,} Error--compression comparison for different
    \(\varepsilon_{\mathrm{RG}}\). The edge-score entropy loss
    \(\left|1-S_{\mathrm{edge}}(\beta)/\log N_e\right|\)
    and specific heat \(C_{\mathrm{edge}}(\beta)\) locate the operating
    points relative to the dominant peak.
    \textbf{d,} Ensemble-averaged bare-Laplacian spectral density,
    ordered nonzero eigenvalues, heat-kernel entropy, and specific heat
    at selected RG levels.
    }
    \label{Fig:S_voronoi_ensemble}
\end{figure*}

Figure~\ref{Fig:S_voronoi_ensemble}a,b shows the case-wise compression and response preservation across the full ensemble. All realizations are coarse-grained using the same ERRG sampling parameter \(\varepsilon_{\mathrm{RG}}=2.8\). At RG level \(\ell=12\), the mean compressed node count is
\[
\left\langle N_{\ell=12}\right\rangle=657.4,
\]
corresponding to a mean node-count ratio
\[
\left\langle\eta_{\ell=12}\right\rangle
=
\left\langle\frac{N_{\ell=12}}{N_0}\right\rangle
=
0.161.
\]
Thus, the coarse node count is approximately \(84\%\) smaller than the original node count. The mean and median shape-response errors are \(10.3\%\) and \(9.9\%\), respectively, while the mean and median shape-cosine mismatches are \(4.15\times10^{-3}\) and \(3.85\times10^{-3}\). The errors remain narrowly distributed across the ensemble, and representative case 06 lies within the main ensemble range. Therefore, at this level of compression, the response preservation shown for case 06 is typical rather than unusually favorable.

Because the RG levels are discrete, the threshold at
\(\ell\approx12.3\) reported in the main text is obtained by interpolation of
the ensemble-median error--compression curve, whereas
Fig.~\ref{Fig:S_voronoi_ensemble}a,b reports the directly evaluated case-wise
statistics at the integer level \(\ell=12\).

The sampling parameter \(\varepsilon_{\mathrm{RG}}\) is selected from the ensemble-level error--compression comparison in Fig.~\ref{Fig:S_voronoi_ensemble}c. Smaller \(\varepsilon_{\mathrm{RG}}\) produces more conservative contraction, whereas larger values compress the networks more rapidly but lead to a faster increase in response error. The intermediate value \(\varepsilon_{\mathrm{RG}}=2.8\) achieves substantial compression while avoiding the more rapid loss of response accuracy observed for the more aggressive choices. It is therefore used for all main text results on multiscale physical networks. The right panel shows the normalized edge-score entropy loss \(1-S_{\mathrm{edge}}(\beta)/\log N_e\) and the corresponding specific heat \(C_{\mathrm{edge}}(\beta)\), evaluated across the original-network ensemble at \(\ell=0\). These quantities locate the tested operating points relative to the dominant reorganization of the ERRG edge-score distribution.

Finally, Fig.~\ref{Fig:S_voronoi_ensemble}d shows the ensemble-averaged evolution of the bare coarse-grained Laplacian statistics at selected RG levels. The mean eigenvalue density \(\langle D(\lambda)\rangle\), mean ordered nonzero eigenvalues \(\langle\lambda_k\rangle\), mean heat-kernel entropy \(\langle S_{\mathrm{HK}}(\tau)\rangle\), and mean heat-kernel specific heat \(\langle C_{\mathrm{HK}}(\tau)\rangle\) retain recognizable overall structures during ERRG, while shifting systematically toward larger eigenvalues and shorter characteristic diffusion times.

\subsection{Central-force mechanical networks with random holes: ensemble robustness and ERRG diagnostics}
\label{SI:delaunay_casewise_level11}

We generate an ensemble of 90 central-force mechanical networks with random holes using the same construction protocol but different random seeds. Each realization starts from exactly \(N_0=4096\) points in a rectangular domain. A subset of points is placed along the boundary, while the remaining points are sampled randomly in the interior. Candidate bonds are constructed by Delaunay triangulation, after which the interior-node positions are smoothed while keeping the boundary nodes fixed. Delaunay edges longer than \(2.5\) times the median edge length are removed both before and after the smoothing step. This long-edge filtering opens randomly distributed pore-like holes across locally sparse regions of the otherwise triangulated network. The coordinates are finally rescaled to a common mean edge length, and all remaining bonds are assigned the same central-force stiffness.

\begin{figure*}[t]
    \centering
    \includegraphics[width=\textwidth]{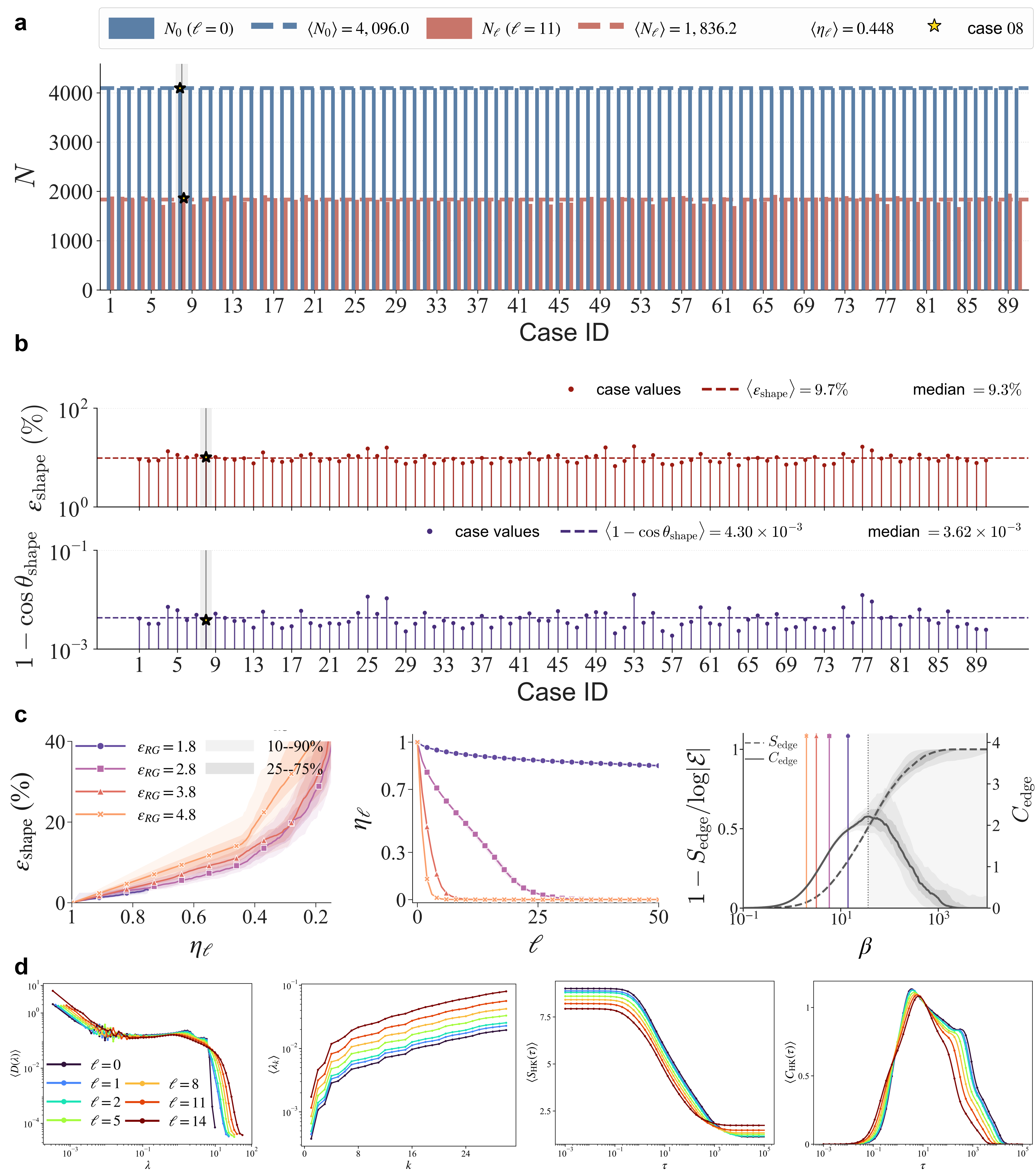}
    \caption{
    \textbf{Ensemble robustness, parameter selection, and spectral
    diagnostics for central-force mechanical networks with random holes.}
    Panels a and b show 90 realizations at \(\ell=11\), using
    \(\varepsilon_{\mathrm{RG}}=2.8\).
    \textbf{a,} Initial and compressed node counts; yellow stars and the
    gray guide identify case 08.
    \textbf{b,} Shape-response error \(\varepsilon_{\mathrm{shape}}\)
    (upper) and shape-cosine mismatch \(1-\cos\theta_{\mathrm{shape}}\)
    (lower), with a shared case-ID axis. Dashed horizontal lines denote
    ensemble means.
    \textbf{c,} Parameter scan for
    \(\varepsilon_{\mathrm{RG}}=1.8,\,2.8,\,3.8,\,4.8\):
    ensemble-median \(\varepsilon_{\mathrm{shape}}\) versus \(\eta_\ell\)
    (left), \(\eta_\ell\) versus \(\ell\) (middle), and
    \(1-S_{\mathrm{edge}}(\beta)/\log N_e\) and
    \(C_{\mathrm{edge}}(\beta)\) (right).
    Shading denotes percentile bands; vertical lines mark the tested
    operating points.
    \textbf{d,} Ensemble-averaged bare-operator spectral density,
    low-order eigenvalues, heat-kernel entropy, and specific heat
    at selected RG levels.
    }
    \label{Fig:S_delaunay_casewise_level11}
\end{figure*}

For each realization, we apply ERRG using the sampling parameter \(\varepsilon_{\mathrm{RG}}=2.8\) and evaluate the same patch-loaded vector response test used in the main text. Figure~\ref{Fig:S_delaunay_casewise_level11} summarizes the ensemble robustness, parameter selection, and spectral diagnostics.

At RG level \(\ell=11\), the ensemble-mean compressed node count is
\[
\left\langle N_{\ell=11}\right\rangle=1836.2,
\]
corresponding to the ensemble-mean node-count ratio
\[
\left\langle\eta_{\ell=11}\right\rangle
=
\left\langle\frac{N_{\ell=11}}{N_0}\right\rangle
=
0.448.
\]
Thus, after 11 ERRG steps, the coarse networks contain, on average, approximately \(45\%\) as many nodes as the original networks.

The vector-response errors remain concentrated near \(10\%\) across the ensemble. The ensemble-mean shape-response error is
\[
\left\langle\varepsilon_{\mathrm{shape}}\right\rangle
=
9.7\%,
\]
with a median value of \(9.3\%\). The corresponding ensemble-mean shape-cosine mismatch is
\[
\left\langle1-\cos\theta_{\mathrm{shape}}\right\rangle
=
4.30\times10^{-3},
\]
with a median value of \(3.62\times10^{-3}\). The small shape-cosine mismatch shows that the lifted coarse responses remain strongly aligned with the fine vector-response fields across the ensemble. Representative case 08 lies within the main ensemble range.

Figure~\ref{Fig:S_delaunay_casewise_level11}c compares the error--compression behavior for different values of \(\varepsilon_{\mathrm{RG}}\). Smaller values produce more conservative coarse graining, whereas larger values reduce the node-count ratio more rapidly but lead to an earlier increase in response error. The intermediate choice \(\varepsilon_{\mathrm{RG}}=2.8\) balances node reduction with response preservation and is therefore used for the main-text results. The normalized edge-score entropy loss and specific heat locate the tested operating points relative to the dominant reorganization of the edge-score distribution.

Figure~\ref{Fig:S_delaunay_casewise_level11}d shows the evolution of the spectral and heat-kernel diagnostics of the bare coarse stiffness operator. The low-frequency spectral structure remains recognizable during the early stages of ERRG, although the eigenvalues evolve systematically as coarse graining proceeds. At stronger coarse graining, the elevated long-\(\tau\) heat-kernel entropy plateau is consistent with the appearance of additional zero or near-zero modes. These results support the main-text conclusion that preservation of the prescribed vector response does not by itself guarantee preservation of mechanical rigidity.

Together, these ensemble results show that representative case 08 reflects the typical response-preservation behavior of the central-force mechanical networks with random holes. At \(\ell=11\), the ensemble-mean node-count ratio is approximately \(0.45\), while the median shape-response error remains below \(10\%\). The accompanying spectral diagnostics also expose the loss of rigidity that limits stronger coarse graining in the present vector implementation.

\subsection{Response preservation across the experimentally characterized networks}
\label{SI:nanowire_casewise_level44}

\subsubsection{Two-dimensional SEM-derived nanowire networks}
\label{2D nanowire networks}

Silver nanowire (AgNW) films were fabricated on glass substrates
by spray-assisted layer-by-layer assembly following the previously
reported procedure~\cite{wu2025layer}.
Branched poly(ethyleneimine) (PEI; \(M_n\approx60{,}000\)) and AgNWs
with a nominal diameter of \(30\,\mathrm{nm}\) and length of
\(50\,\text{\textmu m}\) were obtained from Sigma-Aldrich and Novarials,
respectively. The substrates were washed with ethanol and deionized
water and plasma-cleaned for \(10\,\mathrm{min}\).
An aqueous PEI solution (\(2.5\,\mathrm{mg\,mL^{-1}}\)) and an aqueous
AgNW dispersion (\(0.2\,\mathrm{mg\,mL^{-1}}\)) were sequentially
spray-deposited for \(30\,\mathrm{s}\) and \(100\,\mathrm{s}\),
respectively. AgNW deposition was performed at normal incidence
from a nozzle-to-substrate distance of \(5\,\mathrm{cm}\).
Each deposition was followed by a \(10\,\mathrm{s}\) water-spray
rinse and drying with compressed air. Repeating this sequence
produced PEI/AgNW multilayer films.

Top-view scanning electron microscopy (SEM) images were acquired
at an accelerating voltage of \(2\,\mathrm{kV}\) and a working
distance of \(4.8\,\mathrm{mm}\). Individual images were collected
by mapping and stitched together to obtain the larger fields of
view used for network reconstruction.

Network graphs were extracted from skeletons of binarized SEM
images using StructuralGT~\cite{kadar2024structuralgt}.
Junctions and endpoints of the skeleton define nodes, and the
segments connecting them define edges. For the ERRG tests,
all edges are assigned uniform weights.

We analyze an ensemble of 90 SEM-derived nanowire networks.
For each network, we apply ERRG with the fixed sampling parameter
\(\varepsilon_{\mathrm{RG}}=3.8\) and evaluate the same two-patch
scalar-response test used in the main text.
Case-wise results are reported at RG level \(\ell=44\) to assess
whether the response preservation observed in the representative
network is typical across the experimentally reconstructed ensemble.

\begin{figure*}[t]
    \centering
    \includegraphics[width=\textwidth]{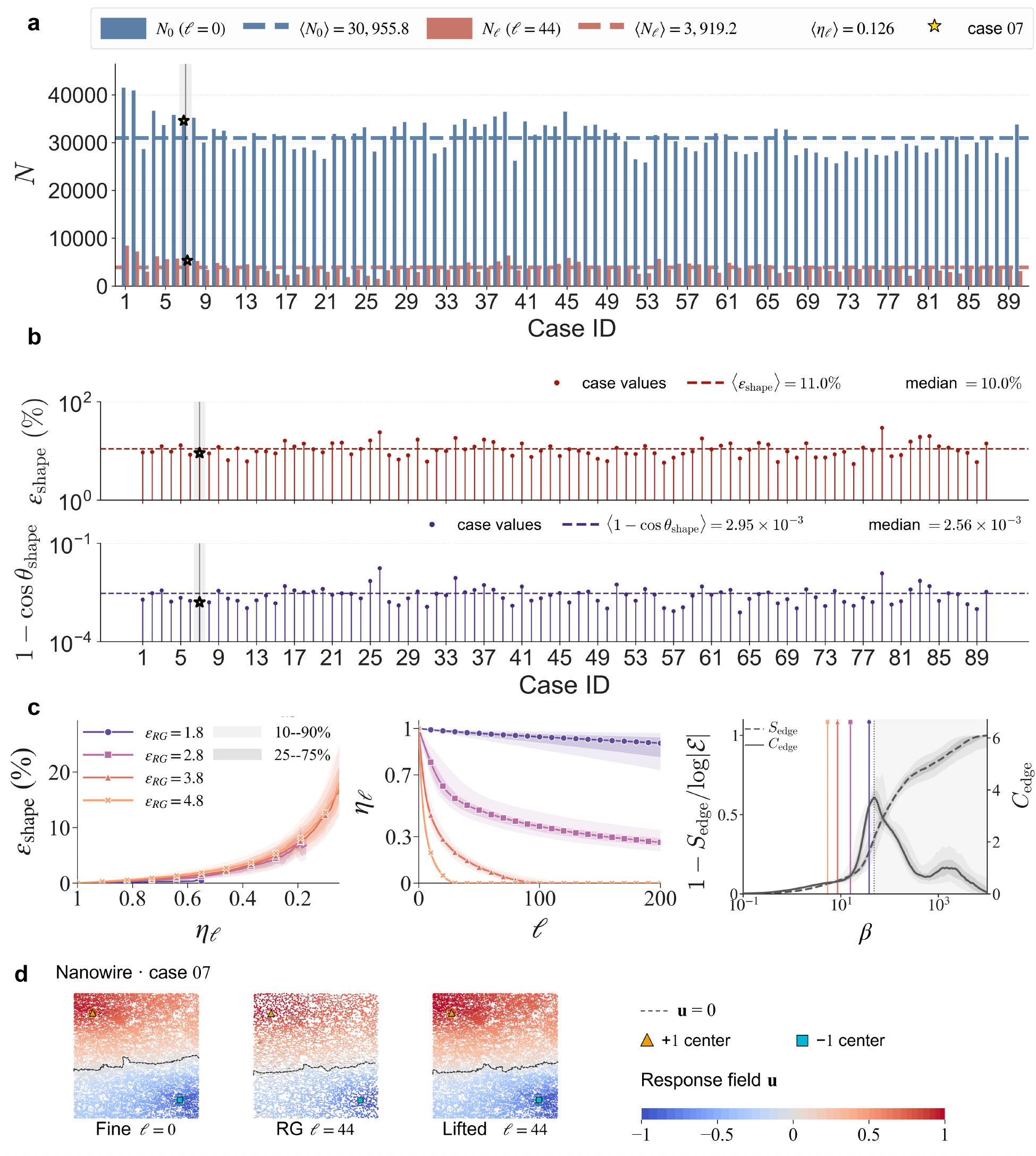}
    \caption{
    \textbf{Response preservation and coarse-graining diagnostics
    for experimentally reconstructed networks.}
    \textbf{a--d,} SEM-derived nanowire ensemble (90 realizations).
    \textbf{a,} Initial and compressed node counts at \(\ell=44\),
    using \(\varepsilon_{\mathrm{RG}}=3.8\).
    \textbf{b,} Case-wise \(\varepsilon_{\mathrm{shape}}\) (upper) and
    \(1-\cos\theta_{\mathrm{shape}}\) (lower) at \(\ell=44\), with a
    shared case-ID axis. Dashed horizontal lines denote ensemble means;
    the yellow star and gray vertical guide identify case 07.
    \textbf{c,} Parameter scan for
    \(\varepsilon_{\mathrm{RG}}=1.8,\,2.8,\,3.8,\,4.8\):
    ensemble-median \(\varepsilon_{\mathrm{shape}}\) versus \(\eta_\ell\)
    (left), \(\eta_\ell\) versus \(\ell\) (middle), and
    \(1-S_{\mathrm{edge}}(\beta)/\log N_e\) and
    \(C_{\mathrm{edge}}(\beta)\) (right).
    Shading denotes percentile bands; vertical lines mark the tested
    operating points.
    \textbf{d,} Two-patch scalar-response fields of SEM case 07
    using \(\varepsilon_{\mathrm{RG}}=3.8\).
    From left to right: fine (\(\ell=0\)), coarse (\(\ell=44\)),
    and lifted coarse (\(\ell=44\)) responses.
    Colors denote \(\mathbf u\); orange triangles and cyan squares
    mark positive and negative loading centers.
    The dashed contour marks \(u=0\).
    }
    \label{Fig:S_nanowire_casewise_level44}
\end{figure*}

Panels~a and b of Fig.~\ref{Fig:S_nanowire_casewise_level44}
summarize the compression and response-preservation results for this
ensemble.
Because the nanowire networks are reconstructed from different SEM images,
their initial graph sizes vary across cases. The ensemble-mean initial
node count is
\[
\langle N_0\rangle=30955.8.
\]
After 44 ERRG steps, the ensemble-mean compressed node count is
\[
\langle N_{\ell=44}\rangle=3919.2.
\]
The corresponding ensemble-mean node-count ratio is
\[
\langle\eta_{\ell=44}\rangle
=
\left\langle\frac{N_{\ell=44}}{N_0}\right\rangle
=
0.126.
\]
Thus, the coarse representations contain, on average, about
\(13\%\) as many nodes as their respective original networks.

The shape-response errors remain concentrated around the \(10\%\) level
across the ensemble. The ensemble-mean shape-response error is
\[
\langle\varepsilon_{\mathrm{shape}}\rangle=11.0\%,
\]
with a median value of \(10.0\%\). This indicates that the dominant spatial
response pattern remains well reproduced after lifting the coarse response
back to the original node space. The shape-cosine mismatch between the
mean-subtracted fine and lifted coarse response fields is
also small:
\[
\left\langle1-\cos\theta_{\mathrm{shape}}\right\rangle
=
2.95\times10^{-3},
\]
with a median value of \(2.56\times10^{-3}\). The small mismatch confirms
that the lifted coarse responses remain strongly aligned with the fine
response fields.

Figure~\ref{Fig:S_nanowire_casewise_level44}c compares the error--compression
behavior for different values of \(\varepsilon_{\mathrm{RG}}\). Increasing
\(\varepsilon_{\mathrm{RG}}\) accelerates the reduction in node-count ratio,
but also causes the response error to increase at earlier RG levels.
The value \(\varepsilon_{\mathrm{RG}}=3.8\) used in the main text is
selected to provide substantial compression while maintaining low response
error. Its operating point lies on the low-\(\beta\) side of the main
\(C_{\mathrm{edge}}(\beta)\) peak, before the strongest change in the
edge-sampling hierarchy.

Figure~\ref{Fig:S_nanowire_casewise_level44}d compares the fine
(\(\ell=0\)), coarse, and lifted coarse (\(\ell=44\)) scalar-response
fields for SEM case 07, using \(\varepsilon_{\mathrm{RG}}=3.8\).

Together, these ensemble results show that the representative SEM-derived
nanowire example reflects typical behavior within the SEM-derived
nanowire ensemble. At \(\ell=44\), the ensemble-mean node-count
ratio is approximately \(0.13\), while the ensemble-median
shape-response error remains at \(10.0\%\) and the shape-cosine
mismatch remains small.

\subsubsection{Three-dimensional ethyl cellulose aerogel networks}
\label{3D aerogel networks}

Ethyl cellulose (EC) aerogels were fabricated from EC nanosheets
by Pickering foam assembly. EC (ethoxyl content of 48\%;
viscosity of \(300\,\mathrm{cP}\) specified for a 5\% solution
in toluene/ethanol, 80:20) and acetone were obtained from
Sigma-Aldrich. EC nanosheets were prepared using the previously
reported shear-driven method~\cite{bang2023fluid}.
A 7 wt\% EC solution in acetone was injected directly into
deionized water in the shear zone of a colloidal mill
(Magic Lab, IKA Works, Germany) equipped with an MK module
operating at \(20{,}000\,\mathrm{rpm}\). The resulting nanosheets
were washed several times with deionized water and redispersed
in water. Pickering foams were formed by mechanical shaking,
drained to remove excess water, and freeze-dried for
\(48\,\mathrm{h}\) to obtain the aerogels.

X-ray micro-computed tomography (micro-CT) was performed using
a Zeiss Versa 520 system (Carl Zeiss X-ray Microscopy,
Pleasanton, CA, USA). EC aerogel samples were cut with a razor
blade into pieces approximately \(1\,\mathrm{cm}\) wide and
placed on the sample holder. Scans were acquired at
\(40\,\mathrm{kV}\) with \(4\times\) optical magnification
and a detector binning factor of 2. For each scan, 2,001
projections were collected over a full \(360^\circ\) rotation,
yielding a voxel size of \(1.2\,\mu\mathrm{m}\).

Tomographic reconstruction and image analysis were performed
in Dragonfly (Comet Group, W\"unnewil-Flamatt, Switzerland).
Upper and lower Otsu thresholds were applied to the
reconstructed volumes to segment the solid and pore phases,
respectively. Three-dimensional network graphs were
reconstructed from the segmented volumes.

\begin{figure*}[t]
    \centering
    \includegraphics[width=\textwidth]{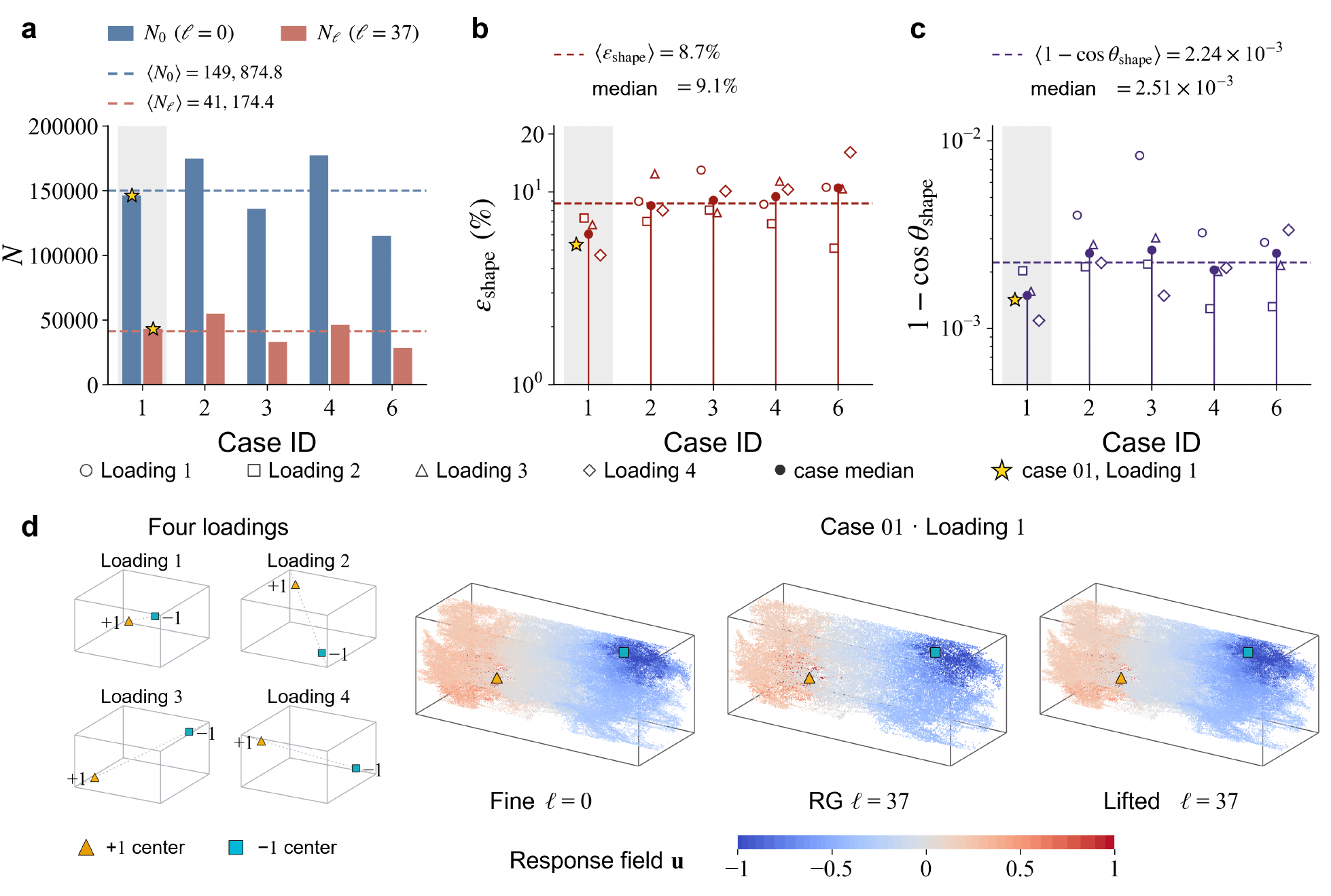}
    \caption{
    \textbf{Response preservation across reconstructed ethyl cellulose
    aerogel networks.}
    \textbf{a,} Initial and compressed node counts for five networks
    at \(\ell=37\). Dashed lines denote ensemble means.
    \textbf{b,c,} Shape-response error \(\varepsilon_{\mathrm{shape}}\)
    (b) and shape-cosine mismatch \(1-\cos\theta_{\mathrm{shape}}\)
    (c) under four two-patch loading configurations per network.
    Open circles, squares, triangles, and diamonds denote loadings
    1--4; filled dots denote within-network medians.
    Dashed horizontal lines denote the corresponding means.
    Yellow stars highlight case 01 and, in (b,c), loading 1.
    \textbf{d,} Four loading configurations (left) and the fine
    (\(\ell=0\)), coarse, and lifted coarse (\(\ell=37\)) response
    fields for case 01 under loading 1 (right).
    Colors denote \(\mathbf u\); orange triangles and cyan squares
    mark positive and negative loading centers.
    Dotted guides connect loading pairs.
    }
    \label{Fig:S_aerogel_casewise_level37}
\end{figure*}

We also evaluate five reconstructed ethyl cellulose aerogel networks, each under four two-patch loading configurations (Fig.~\ref{Fig:S_aerogel_casewise_level37}a--c).
At \(\ell=37\), the mean node count decreases from
\(149874.8\) to \(41174.4\).
Figure~\ref{Fig:S_aerogel_casewise_level37}d shows the 4 loading configurations and the fine, coarse, and lifted coarse response fields for case 01 under loading 1.
Figure~\ref{Fig. 4}j shows the error--compression tradeoff for the five aerogel networks.

\subsubsection{Common procedures}
\label{Common procedures}

For both the SEM-derived nanowire ensemble and the representative
three-dimensional ethyl cellulose aerogel network, edge scores are estimated
using Johnson--Lindenstrauss random projections
\cite{spielman2008graph,spielman2004nearly}.
With the projection parameter \(\varepsilon_{\mathrm{JL}}=0.30\),
the numerator and denominator of \(\chi_e\) are each estimated at
every RG level using
\(16\log N_e/\varepsilon_{\mathrm{JL}}^2\) probes, rounded up
to an integer and processed in blocks of at most \(128\), where
\(N_e\) is the current edge count.
The shifted linear solves use \(\alpha_{\mathrm{leak}}=10^{-12}\) for the SEM networks
and \(\alpha_{\mathrm{leak}}=10^{-6}\) for the three-dimensional network.

\subsection{Comparing ERRG with other network renormalization methods}
\label{SI:nanowire_method_comparison}

\begin{figure*}[t]
    \centering
    \includegraphics[width=\textwidth]{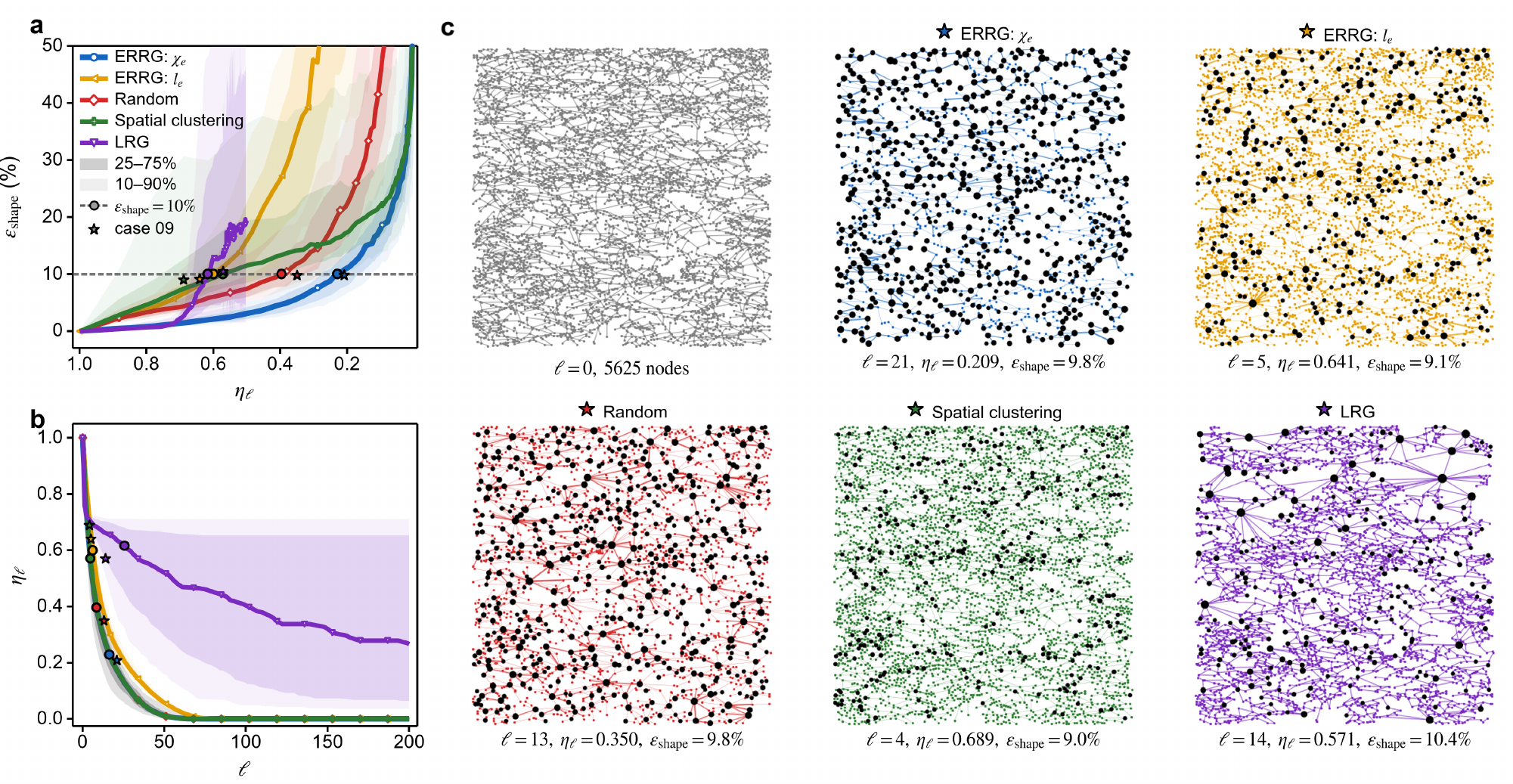}
    \caption{
    \textbf{Comparison between ERRG and other network RG methods using SEM-derived nanowire networks as an example.}
    \textbf{a,} Lifted shape-response error
    \(\varepsilon_{\mathrm{shape}}\) versus node-count ratio
    \(\eta_\ell\) for ERRG based on \(\chi_e\), ERRG based on
    \(l_e\), random contraction, spatial clustering, and LRG.
    \textbf{b,} Node-count ratio \(\eta_\ell\) versus RG level
    \(\ell\) for the same set of methods in (a).
    \textbf{c,} Representative coarse networks at levels nearest
    to \(10\%\) shape-response error.
    For similar initial compression rates, we use
    \(\varepsilon_{\mathrm{RG}}=3.8\) for \(\chi_e\)-based ERRG,
    \(\varepsilon_{\mathrm{RG}}=5.8\) for \(l_e\)-based ERRG,
    and \(\tau_{\mathrm{LRG}}=0.9\) for LRG.
    }
    \label{Fig:method_comparison}
\end{figure*}

Here we compare ERRG with other network RG methods, in terms of preservation of physical responses,  
using nanowire networks as an example system. 

We construct an ensemble of 90 smaller nanowire networks by randomly
cropping equally sized regions from the SEM nanowire images and
applying the same image-to-graph workflow used for the larger
experimental networks in the main text. Self-loops and small
disconnected components are removed, and the largest connected
component is retained. Because the nanowire density varies between
cropped regions, the initial network sizes differ across cases, with
an ensemble-mean node count
\[
\langle N_0\rangle=4779.4.
\]
These smaller networks make the comparison with the dense
implementation of LRG computationally tractable.

For each network, we compare ERRG based on the edge-weight
regularized response time \(\chi_e\), ERRG based on the edge leverage
score \(l_e\), random contraction, spatial clustering, and LRG.
The random baseline uses the same omitted-edge metagraph,
connected-component block construction, and Kadanoff supernode
projection as \(\chi_e\)-based ERRG, but replaces response-weighted
edge sampling with uniform random edge selection. At every RG
level, its target node count is matched to the \(\chi_e\)-based
ERRG trajectory. The spatial baseline selects cluster centers by
Euclidean farthest-point sampling and assigns each node to its
nearest center, using the same target node-count trajectory.
LRG constructs coarse blocks from the heat-kernel density matrix
following the Laplacian renormalization framework
\cite{villegas2023laplacian}.

To obtain comparable initial compression rates, we use
\(\varepsilon_{\mathrm{RG}}=3.8\) for \(\chi_e\)-based ERRG,
\(\varepsilon_{\mathrm{RG}}=5.8\) for \(l_e\)-based ERRG,
and \(\tau_{\mathrm{LRG}}=0.9\) for LRG.
The trajectories are followed for up to 200 RG levels.
All methods are evaluated using the same two-patch scalar-response
test. For each method, the operating level is determined from the
ensemble-median error--compression trajectory near
\(\varepsilon_{\mathrm{shape}}=10\%\), and the case-wise statistics
are evaluated at the corresponding integer RG level.

Overall, \(\chi_e\)-based ERRG achieves the lowest shape-response
error, and the advantage is clearest at deep compression
(Fig.~\ref{Fig:method_comparison}a).
The comparison with random contraction isolates the effect of
response-weighted edge sampling under the same target node-count
trajectory and block-construction procedure.
Similarly, the comparison with spatial clustering highlights that
the edge score is not simply controlled by Euclidean distances,
but encodes physics-specific distances in the network.

The comparison between \(l_e\) and \(\chi_e\) further shows that
incorporating the slow-response time scale produces more effective
coarse variables than using the static edge response alone.
For the parameter choice used here, LRG performs well at shallow
compression (\(\eta_\ell\gtrsim0.7\)), but shows larger
shape-response errors and slower node-count reduction across RG
steps at stronger compression, accompanied by the formation of
large hubs (Fig.~\ref{Fig:method_comparison}a--c).

\begin{figure*}[t]
    \centering
    \includegraphics[width=.9\textwidth]{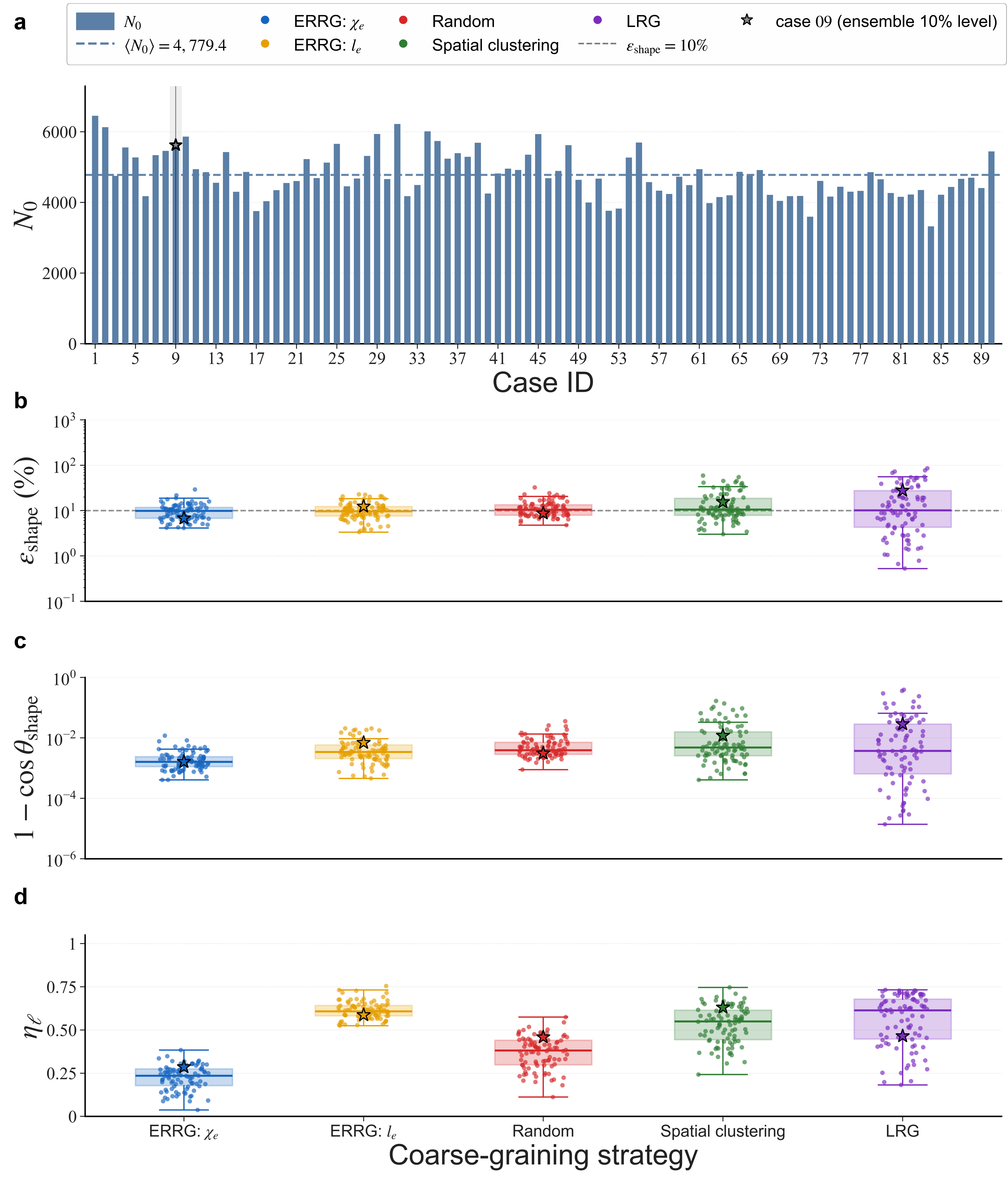}
    \caption{
    \textbf{Case-wise comparison of network RG methods on
    cropped SEM-derived nanowire networks.}
    \textbf{a,} Initial node count \(N_0\) for the 90 cropped
    networks. The dashed line denotes the ensemble mean.
    The star marks the representative network shown in
    Fig.~\ref{Fig:method_comparison}c.
    \textbf{b,} Case-wise shape-response error across the five
    coarse-graining strategies. Each method is evaluated at its
    own ensemble operating level nearest to an ensemble-median
    shape-response error of \(10\%\). The dashed line marks
    \(\varepsilon_{\mathrm{shape}}=10\%\).
    \textbf{c,} Corresponding shape-cosine mismatch
    \(1-\cos\theta_{\mathrm{shape}}\).
    \textbf{d,} Corresponding node-count ratio.
    Dots denote individual reconstructed networks, boxes show the
    \(25\text{--}75\%\) interquartile range, horizontal lines
    inside the boxes denote medians, and stars mark the
    representative network.
    Method colors are consistent with
    Fig.~\ref{Fig:method_comparison}.
    }
    \label{Fig:S_nanowire_method_comparison}
\end{figure*}

Figure~\ref{Fig:S_nanowire_method_comparison} summarizes the
case-wise variability across the same cropped-network ensemble.
Although the methods are evaluated at operating levels near the
same ensemble-median shape-response error, they produce
substantially different node-count ratios.
In particular, \(\chi_e\)-based ERRG reaches the smallest median
\(\eta_\ell\), showing that its compression advantage is not
limited to the representative network.

Together, these results show that incorporating edge-response
information improves preservation of the prescribed physical
response across the cropped SEM-derived ensemble.
They highlight the central idea of ERRG: the relevant degrees
of freedom are defined neither purely in real space nor solely
by the free modes of the network, but by their contribution to
the physical response that the reduced model is intended
to preserve.

\subsection{Temporal response and time-scale renormalization}
\label{SI:temporal_additional_diagnostics}

\begin{figure*}[t]
    \centering
    \includegraphics[width=\textwidth]{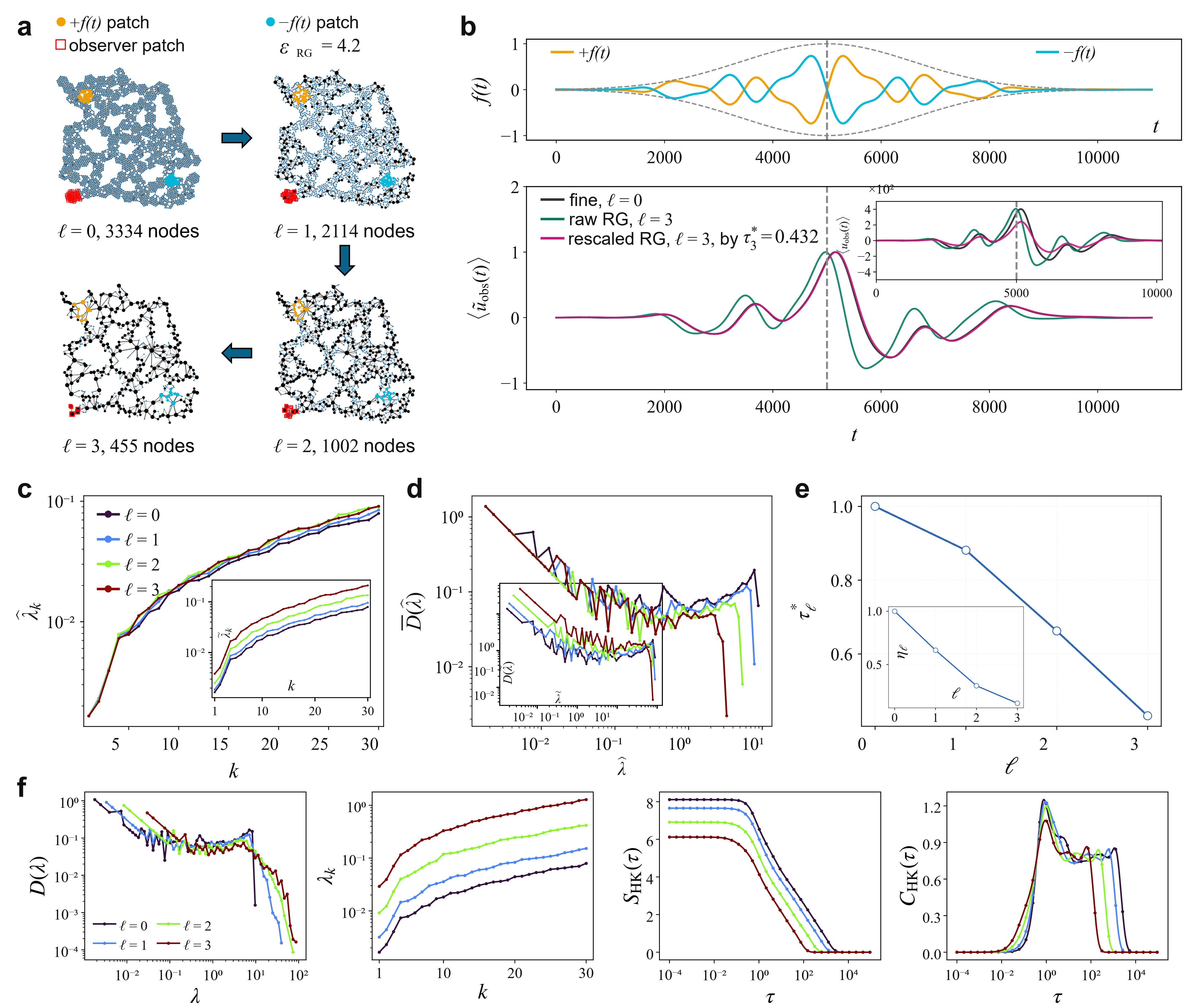}
    \caption{
    \textbf{ERRG preserves temporal response through time-scale
    renormalization.}
    \textbf{a,} ERRG trajectory and temporal-response setup.
    Orange and cyan mark the positive and negative loading patches,
    respectively; red squares mark the observer patch.
    \textbf{b,} Input signals (upper) and normalized observer responses
    (lower), comparing the fine network with the raw and time-rescaled
    coarse network. The inset shows unnormalized responses.
    Dashed curves indicate the Gaussian envelope, and vertical dashed
    lines mark the pulse center.
    \textbf{c--e,} Rescaled low-order eigenvalues (c),
    mode-count-corrected spectral density (d), and time-rescaling
    factor (e). The corresponding unrescaled spectra and node-count
    ratio are shown in the insets.
    \textbf{f,} Bare-Laplacian spectral density, low-order eigenvalues,
    heat-kernel entropy, and specific heat, from left to right,
    without drag normalization or time rescaling.
    The fixed sampling parameter is \(\varepsilon_{\mathrm{RG}}=4.2\).
    }
    \label{Fig:voronoi_temporal_response}
\end{figure*}

We test whether ERRG preserves the temporal response of a multiscale physical network of the same type as those used in
Fig.~\ref{Fig. 2}. With the fixed sampling parameter
\(\varepsilon_{\mathrm{RG}}=4.2\), the node counts decrease from
\(3334\) to \(2114\), \(1002\), and \(455\) at
\(\ell=0,1,2,3\), respectively
(Fig.~\ref{Fig:voronoi_temporal_response}a).

Opposite time-dependent inputs \(\pm f(t)\) are applied to two
distant patches, and the response is measured at a third, spatially
separated observer patch. The input is
\[
f(t)=e^{-(t-t_0)^2/(2\sigma^2)}
\sum_{j=1}^{3}A_j\sin[\omega_j(t-t_0)],
\]
with \(t_0=5000\), \(\sigma=1600\), and
\((A_1,A_2,A_3)=(0.3,0.5,0.2)\).
The driving frequencies are \(\omega_j=2\pi n_j/T_{\max}\), where
\(T_{\max}=11000\) and \((n_1,n_2,n_3)=(3,7,14)\), giving
\[
(\omega_1,\omega_2,\omega_3)
\approx(1.039,2.424,4.847)\widetilde{\lambda}_1^{(0)}.
\]

The fine and coarse networks evolve according to
Eq.~\eqref{Eq:response_real_time}, with the corresponding operator
\(\mathbf L^{(\ell)}\) and accumulated drag matrix
\(\boldsymbol{\Gamma}^{(\ell)}\).
For this overdamped test, the node-weight matrix in
Eq.~\eqref{EQ:drag_normal} is
\(\boldsymbol{W}^{(\ell)}=\boldsymbol{\Gamma}^{(\ell)}\).
Following Sec.~\ref{SI:spectral_heat_kernel}, we use
\(\tau_\ell^*=\widetilde{\lambda}_1^{(0)}/
\widetilde{\lambda}_1^{(\ell)}\), where
\(\widetilde{\lambda}_1^{(\ell)}\) is the first nonzero eigenvalue of
\(\widetilde{\mathbf L}^{(\ell)}\).
The time-rescaled response is obtained from
\begin{equation}
\left[
\frac{1}{\tau_\ell^*}
\boldsymbol{\Gamma}^{(\ell)}\partial_t
+
\mathbf L^{(\ell)}
\right]
\mathbf u^{(\ell)}(t)
=
\mathbf f^{(\ell)}(t).
\end{equation}
The raw RG response uses the same equation without the factor
\(1/\tau_\ell^*\).

Figure~\ref{Fig:voronoi_temporal_response}b compares the
fine response at \(\ell=0\) with the raw and time-rescaled coarse
responses at \(\ell=3\).
The spectral rescaling gives \(\tau_3^*=0.432\), without fitting the
time-domain signal. Before rescaling, the coarse response retains the
dominant waveform but exhibits a temporal mismatch. Applying
\(\tau_3^*\) improves its temporal alignment with the fine response,
while the unnormalized responses in the inset show that some amplitude
differences remain. This provides a time-domain check of the spectral
rescaling for the prescribed signal after substantial network reduction.

Panels c and d show the rescaled low-order eigenvalues
\(\widehat{\lambda}_k^{(\ell)}=
\tau_\ell^*\widetilde{\lambda}_k^{(\ell)}\)
and the mode-count-corrected spectral density, with the corresponding
unrescaled drag-normalized quantities in the insets. Panel e shows
\(\tau_\ell^*\) and the node-count ratio \(\eta_\ell\).
For comparison, panel f shows the spectral and heat-kernel diagnostics
of the bare coarse Laplacian \(\mathbf L^{(\ell)}\), using the definitions
in SI Sec.~\ref{SI:spectral_heat_kernel}.
These bare-operator diagnostics use the raw diffusion time \(\tau\),
without drag normalization or time rescaling, and show the uncorrected
spectral flow generated by ERRG.

\subsection{Preservation of classification decisions under ERRG}
\label{sec:errg-task-preservation}

\begin{figure*}[t]
    \centering
    \includegraphics[width=\textwidth]{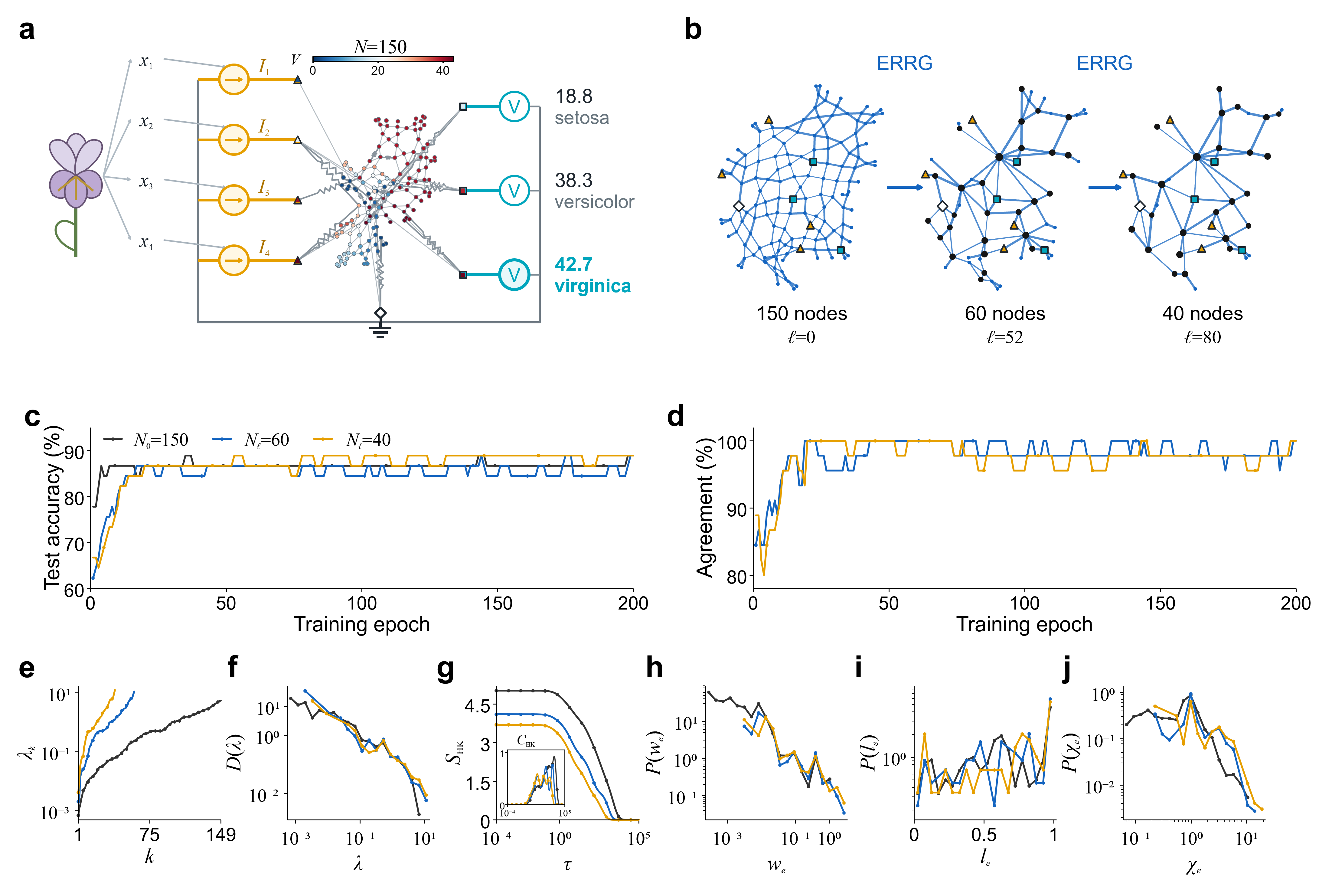}
    \caption{
    \textbf{ERRG flow of a trained electrical classifier.}
    \textbf{a,} Iris classification with 4 current inputs and 3
    voltage outputs. The largest output voltage selects the class.
    Node colors show the voltage relative to ground. Gold triangles,
    cyan squares and an open diamond mark the inputs, outputs and
    ground, respectively.
    \textbf{b,} Network configurations at RG levels $\ell=0,52,80$,
    with 150, 60 and 40 nodes, respectively
    ($\varepsilon_{\mathrm{RG}}=2.4$).
    \textbf{c,d,} Test accuracy (\textbf{c}) and prediction agreement
    with the 150-node network (\textbf{d}) on the same 45 test samples
    at each of the 200 saved training epochs. The node groups in
    \textbf{b} are fixed, and the conductances are taken from the
    corresponding epoch.
    \textbf{e--j,} Network measures at the final training epoch.
    Ordered positive eigenvalues (\textbf{e}), spectral density
    (\textbf{f}), heat-kernel entropy and specific heat
    (\textbf{g}, main panel and inset), and distributions of edge
    weights $w_e$ (\textbf{h}), leverage scores $l_e$ (\textbf{i})
    and edge-weight-regularized response times $\chi_e$ (\textbf{j}).
    Colors in \textbf{d--j} follow the legend in \textbf{c}.
    }
    \label{fig:errg-task-preservation}
\end{figure*}

To explore whether ERRG preserves neural network functions across scales, we apply ERRG to a 150-node Mikado electrical resistor network trained for Iris classification using the method of Ref.~\cite{li2024training}.

Using $\chi_e$ as the edge score, we start from the network at the end of training and apply successive ERRG steps, keeping the input, output, and ground nodes separate. Along this flow, we examine the 60-node and 40-node networks reached at RG levels $\ell=52$ and $80$ (Fig.~\ref{fig:errg-task-preservation}a,b). For each of these two networks, we record which original nodes have been merged into each supernode.

We then use these same node groups to coarse-grain the networks saved during training. For example, original nodes that belong to the same supernode in the 60-node network are kept together when we go back to any earlier saved epoch. At that epoch, we calculate the conductance between two supernodes by summing the conductances of all original edges connecting their two groups, using the weights saved at that epoch. We repeat this procedure for both the 60-node and 40-node groups across all 200 saved training epochs. The nodes belonging to each supernode remain fixed, while the conductances between supernodes change as the original edge weights evolve during training.

On the reconstructed test set of 45 samples, test accuracy measures the percentage of samples that each network classifies correctly. At the final saved epoch, all three networks correctly classify 40 of the 45 samples, giving the same test accuracy of 88.9\% (Fig.~\ref{fig:errg-task-preservation}c).

Prediction agreement measures the percentage of samples for which a reduced network predicts the same class as the original 150-node network at the same saved epoch. 
Across the 200 saved training epochs, this agreement averages 98.0\% for the 60-node network and 97.5\% for the 40-node network. 
At the final saved epoch, both reduced networks match all 45 predictions of the original network, giving 100\% agreement (Fig.~\ref{fig:errg-task-preservation}d). 
These results show that, for this network and reconstructed test set, the node groups identified at the end of training retain most classification decisions throughout training, even after a 73.3\% reduction in node count.

We compare the original 150-node network with its 60-node and 40-node ERRG reductions at the same final trained state (Fig.~\ref{fig:errg-task-preservation}e--j).
The Laplacian spectrum and edge distributions change with compression, while all three networks give the same predictions on the 45 test samples.

\end{document}